\documentclass[10pt,compsoc]{IEEEtran}

\usepackage{verbatim}
\usepackage{url}
\usepackage{bm}
\usepackage{color}
\usepackage{balance}
\makeatletter
\newif\if@restonecol
\makeatother

\usepackage[lined,algonl,boxed]{algorithm2e}
\usepackage{epsfig}
\usepackage{epstopdf}
\usepackage{graphicx}
\usepackage{subfigure}
\usepackage{multirow}
\usepackage{amsmath}
\usepackage{amssymb}

\ifCLASSOPTIONcompsoc
\else
\fi

\ifCLASSINFOpdf
\else
\fi

\newcommand{\hide}[1]{} 
\newcommand{\vpara}[1]{\vspace{0.03in}\noindent\textit{#1 }}

\newcommand{\beq}[1]{\vspace{0.0in}\begin{equation}#1\end{equation}\vspace{-0.01in}}

\newtheorem{definition}{Definition}
\newtheorem{problem}{Problem}

\def\hindex{\textit{h}-index}
\def\hindices{\textit{h}-indices}
\def\fonescore{$F_1$ score}

\begin{document}

\title{Can Scientific Impact Be Predicted?}

\author{
Yuxiao Dong$^{\ddag}$, Reid A. Johnson$^{\ddag}$, Nitesh V. Chawla
\IEEEcompsocitemizethanks{
\IEEEcompsocthanksitem 
The authors are with the Department of Computer Science and Engineering and the Interdisciplinary Center for Network Science and Applications (iCeNSA), University of Notre Dame, IN 46556.
\protect \\
E-mail: \{ydong1, rjohns15, nchawla\}@nd.edu \protect \\
$^{\ddag}$The first two student authors contributed equally to this work. \protect \\
To whom correspondence should be addressed. E-mail: nchawla@nd.edu
}
}

\markboth{IEEE Transactions on Big Data, ~Vol.~2, No.~1, 2016}{Dong \MakeLowercase{\textit{et al.}}: Can Scientific Impact Be Predicted?}

\IEEEcompsoctitleabstractindextext{%

\begin{abstract}

A widely used measure of scientific impact is citations. However, due to their heavy-tailed distribution, citations are fundamentally difficult to predict. Instead, to characterize scientific impact, we address two analogous questions asked by many scientific researchers: ``How will my \hindex\ evolve over time, and which of my previously or newly published papers will contribute to it?'' To answer these questions, we perform two related tasks. First, we develop a model to predict authors' future \hindices\ based on their current scientific impact. Second, we examine the factors that drive papers---either previously or newly published---to increase their authors' predicted future \hindices. By leveraging relevant factors, we can predict an author's \hindex\ in five years with an $R^2$ value of 0.92 and whether a previously (newly) published paper will contribute to this future \hindex\ with an \fonescore\ of 0.99 (0.77). We find that topical authority and publication venue are crucial to these effective predictions, while topic popularity is surprisingly inconsequential. Further, we develop an online tool that allows users to generate informed \hindex\ predictions. Our work demonstrates the predictability of scientific impact, and can help scholars to effectively leverage their position of ``standing on the shoulders of giants.''

\end{abstract}

}

\maketitle
\IEEEdisplaynotcompsoctitleabstractindextext

\sloppy

\IEEEraisesectionheading{\section{Introduction}
\label{sec:intro}
}

\IEEEPARstart{S}{cientific} impact plays a pivotal role in the evaluation of the output of scholars, departments, and institutions. 
Scientific researchers generate scientific impact through novel discoveries and developments, which are traditionally disseminated to a wider community via publications. The impact of each of these findings and corresponding publications---both to a field of research and, by extension, to the reputation of the author---can be affected by a variety of factors, which may be directly or indirectly related to the findings themselves.  
Due to the confluence of such factors, a researcher's body of work is likely to be composed of findings and publications of varying impact. 
Consequently, it can be challenging to predict a researcher's future impact and the influence of any particular publication on this impact, regardless of how impact is measured.

Often a researcher's total number of citations is used as a measure of impact, while a researcher's total number of publications is used as a measure of productivity. However, while these simple measures are intuitive and can be useful, they also have significant limitations. For example, a solitary well-cited, impactful paper can skew the total number of citations, potentially distorting its use as a measure of overall impact. Similarly, the total number of publications can be increased by a large number of poorly cited papers, which may not be indicative of the actual productivity involved. Moreover, as citations demonstrate a heavy-tailed distribution, with the vast majority of publications receiving few citations, these simple measures are exceedingly difficult to estimate using traditional regression analysis \cite{Radicchi:PNAS08,Cheng:WWW14}. Thus, determining how many citations a given researcher or a given paper will receive is often ineffective in practice.

In light of these difficulties and limitations, we instead address two analogous questions asked by many academic researchers: ``\textit{\textbf{How will my \hindex\ evolve over time, and which of my previously and newly published papers will contribute to my future \hindex?}}''

These questions are based on the \hindex. As described by J. E. Hirsch, by whom the index was proposed: ``A scientist has index $h$ if $h$ of his or her papers have at least $h$ citations each, and the other papers have no more than $h$ citations each'' \cite{Hirsch:05}.  
The \hindex\ is thus a function of the number of publications (quantity) and the number of citations per publication (quality). As a result of its simplicity and predictive value, the \hindex\ has become a \textit{de facto} standard for measuring scientific impact.

\begin{figure}[t]
\centering
\includegraphics[width=3.2in]{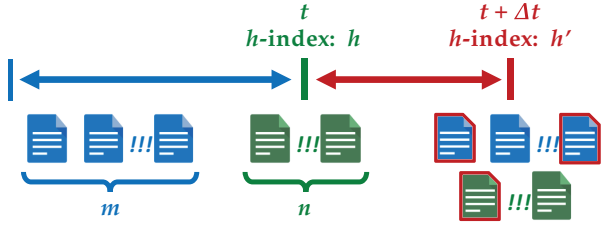}

\caption{\label{fig:intro-exp}
{\bf Illustrative example of scientific impact prediction.}
{
Before time $t$, a scholar published $m$ papers and had an \hindex\ of $h$. 
Our prediction problems are targeted at answering two questions: 
1) What is the scholar's future \hindex, $h'$, at time $t$+$\Delta t$? 
2) Which of his/her papers, both (a) those $m$ papers previously published before $t$ and (b) those $n$ new papers published at $t$, will contribute to $h'$?
}
}
\vspace{-0.1cm}
\end{figure}

\vpara{Present Work.} 
To tackle the questions of how one's \hindex\ will evolve over time and which publications will contribute to it, we formulate two scientific impact prediction problems, as shown in Figure \ref{fig:intro-exp}. 
Our first task is to predict authors' future \hindices\ based on their current scientific impact, which has been explored with data on a small sample of neuroscientists~\cite{Acuna:Nature12}. 
We then determine whether a given paper will influence a particular author's predicted future \hindex, which we formalize as our primary scientific impact prediction problem. 
Accordingly, our second (primary) prediction problem is to determine whether a given previously or newly published paper will, after a predefined timeframe, increase the \textit{future} \hindex\ of its primary author (i.e., the paper's first author or the author with the highest \hindex).  
The predicted future \hindices\ generated by the first task are used as the future \hindices\ in our primary task. 
Thus, in our primary task, an author's future \hindex\ represents the author's expected \hindex\ after the predefined period of time, with the purpose of accounting for the change in the author's \hindex\ over the prediction timeframe.

\vpara{Contributions.} 
This work expands on our previous work~\cite{Dong:WSDM2015}, which aims to discern the impact of a given publication on the primary author's \hindex, in several ways. 
First, we investigate the factors that influence the development of an author's scientific impact, for which we generate a model to infer an author's future \hindex. 
Second, by using the future \hindex\ predicted by this model as the target variable for predicting whether a paper will increase its primary author's \hindex, we account for the dynamic change in the primary author's \hindex\ over the course of prediction timeframe. 
In other words, in this work we aim to predict not only on the newly published papers~\cite{Dong:WSDM2015}, but also on the previously published ones. 
We also re-define the primary author of a publication as both the first author and the author with highest \hindex\ among the author list. 
To further add to the utility of this work, we have also developed and deployed an online tool that allows users to generate \hindex\ predictions based on our findings.

\vpara{Challenges.} 
Factors such as the researcher's current influence, the publication topic, and the publication venue may, among many other factors, play a role in determining the degree to which a publication contributes to the researcher's future impact. 
A resulting challenge is the interplay of such factors, which can confound attempts to generate effective predictions. 
Considerations such as the variability of the \hindex\ according to the ``academic age'' of a researcher, the widely differing citation conventions among different fields, and the co-authorship of researchers with differing \hindices\ can make it difficult to isolate the degree to which a given paper will contribute to the measured impact of its authors. 
Further, effectively predicting whether a publication will contribute to its authors' measured \textit{future} impact must account for the change in impact over the prediction timeframe, which may follow a trajectory and rate particular to each author. 
Our work focuses on addressing and overcoming each of these issues to generate novel, effective scientific impact predictions, as well as investigating precisely what role a variety of factors play in these predictions.

\begin{figure}[t]
\centering
\hspace{-0.25in}
\subfigure[\scriptsize \hindex]{
\label{figsub:intro-r2}
\includegraphics[width=1.2in]{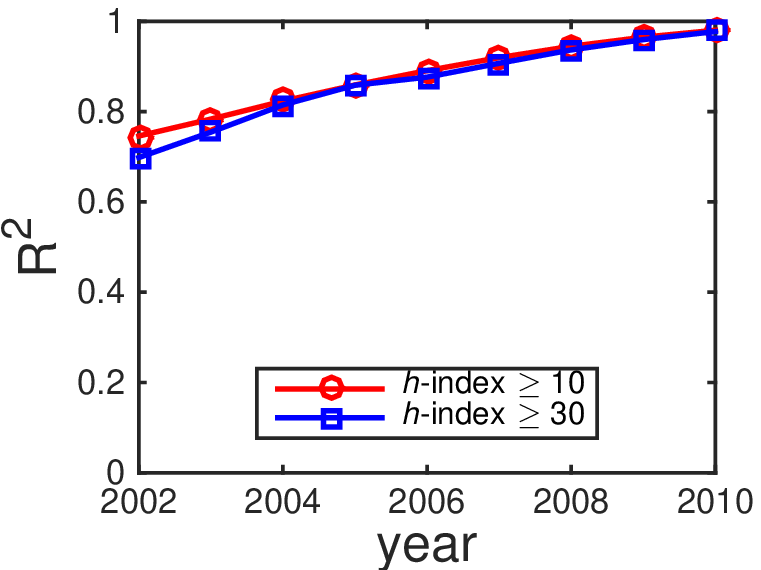}
}
\hspace{-0.25in}
\subfigure[\scriptsize $P^{max}$]{
\label{figsub:intro-f1-max}
\includegraphics[width=1.2in]{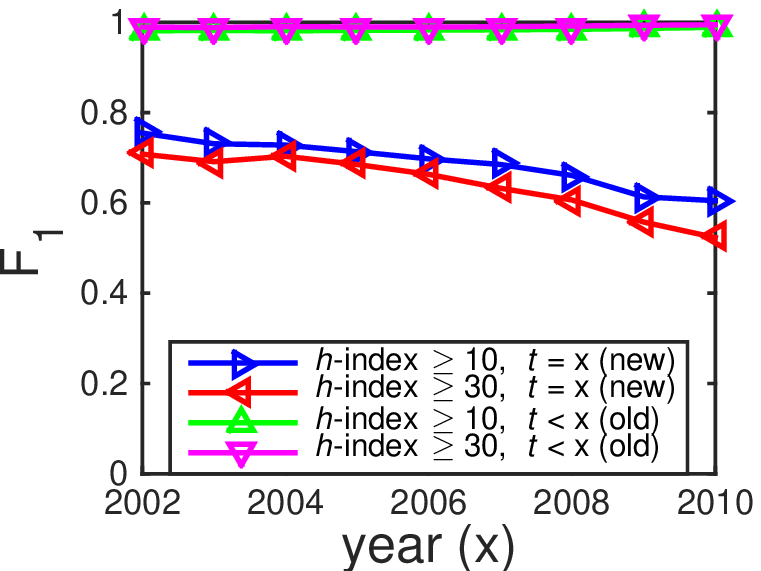}
}
\hspace{-0.2in}
\subfigure[\scriptsize $P^{first}$]{
\label{figsub:intro-f1-first}
\includegraphics[width=1.2in]{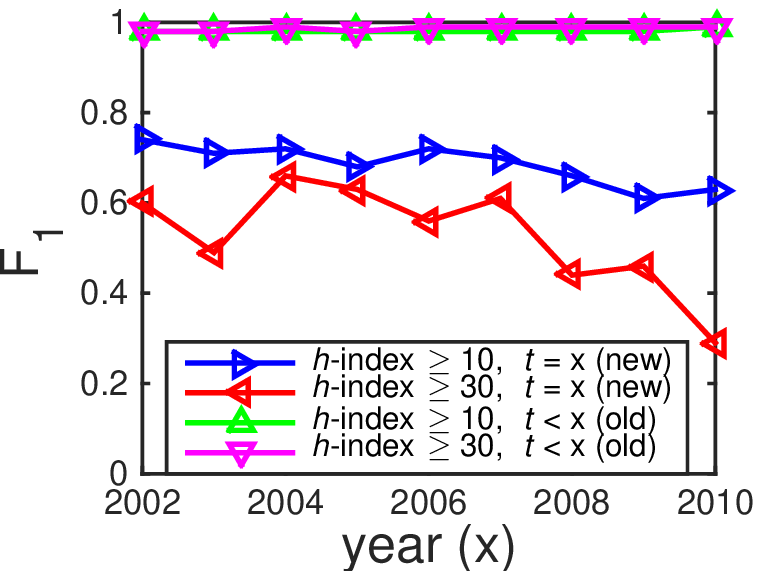}
}
\caption{\label{fig:intro-f1}
{\bf Predictability of scientific impact.} 
{
$x$-axis: year of data used to predict to 2012.
$y$-axis: performance.
(a) Performance for predicting an author's \hindex\ as a regression task ($R^2$ value). 
(b) Performance for predicting whether a given paper will increase the \hindex\ of its primary author (as defined by the author with highest \hindex\ among its author list) as a classification task ($F_1$ score). 
(c) Performance for predicting whether a paper will increase the first author's \hindex.  
}
}
\vspace{-0.1cm}
\end{figure}

\vpara{Results.}
We demonstrate a high level of predictability for scientific impact as measured by our two problems. 
Accordingly, we find strong performance for our first task of predicting an author's future \hindex. Our results demonstrate that we can predict an author's \hindex\ in five years with an $R^2$ value of 0.9197, as shown in Figure~\ref{figsub:intro-r2}. 
This performance generally increases as the prediction timeframe is shortened, with a prediction of ten years achieving an $R^2$ of 0.7461. 
We also find strong performance for our primary task of predicting whether a publication will contribute to its primary author's future \hindex. 
Our results demonstrate that we can predict whether in five years a previously (newly) published paper will contribute to the future \hindex\ of the author with highest \hindex\ with an \fonescore\ of 0.99 (0.77), as shown in Figure~\ref{figsub:intro-f1-max}, an improvement of +130\% (+160\%) over random guessing. 
From Figure~\ref{figsub:intro-f1-first}, we can observe that similar, strong performance is achieved when considering the first author of a publication as its primary author. 
Predictive performance for newly published papers generally increases as the prediction timeframe is expanded. 
However, predictive performance for previously published papers achieves consistently high \fonescore s, suggesting their general predictability. 
Our results also indicate that authors with low \hindices\ are easier to predict for than those with high ones (see Figures~\ref{figsub:intro-f1-max} and \ref{figsub:intro-f1-first}, blue vs. red lines).  

We also assess the influence of various factors on our predictive results.
For our first problem, predicting an author's future \hindex, we find that the author's current \hindex\ is the most important, followed by the number of publications and co-authors. 
For our primary problem, predicting whether a paper will contribute to its primary author's \hindex, we find that topical authority is the most telling factor for newly published papers, while the existing citation information is the most telling for previously published ones, followed by the authors' influence and the publication venue. 
We also find that the venue in which the paper is published and the author's collaborations are moderately significant factors over longer prediction periods, but become inconsequential for shorter ones. 
Finally, we are surprised to find that the popularity of the publication topic has no discernible correlation to the prediction target for both previously and newly published papers. 
Overall, our findings unveil the predictability of scientific impact and provide researchers with concrete suggestions for expanding their scientific influence and, ultimately, for more effectively ``standing on the shoulders of giants.''

\vpara{Data.} In this paper, we use the real-world academic dataset\footnote{\scriptsize The dataset is publicly available at \url{https://aminer.org/billboard/citation} and \url{https://aminer.org/billboard/AMinerNetwork}.} from ArnetMiner \cite{Tang:08KDD}, which is the world-leading free online service for academic social network analysis and mining.
The dataset contains 1,712,433 authors with 2,092,356 papers from computer science venues held until 2012. 
Each paper includes information on the title, abstract, authorship, references, and publication venue and year.
The dataset also captures 4,258,615 collaboration (co-authorship) relationships and 8,024,869 citation relationships. 

\begin{figure}[t]
\centering
\subfigure[\scriptsize Distribution of citations counts]{
\hspace{-0.2in}
\label{figsub:citation-powerlaw}
\includegraphics[width=1.75in]{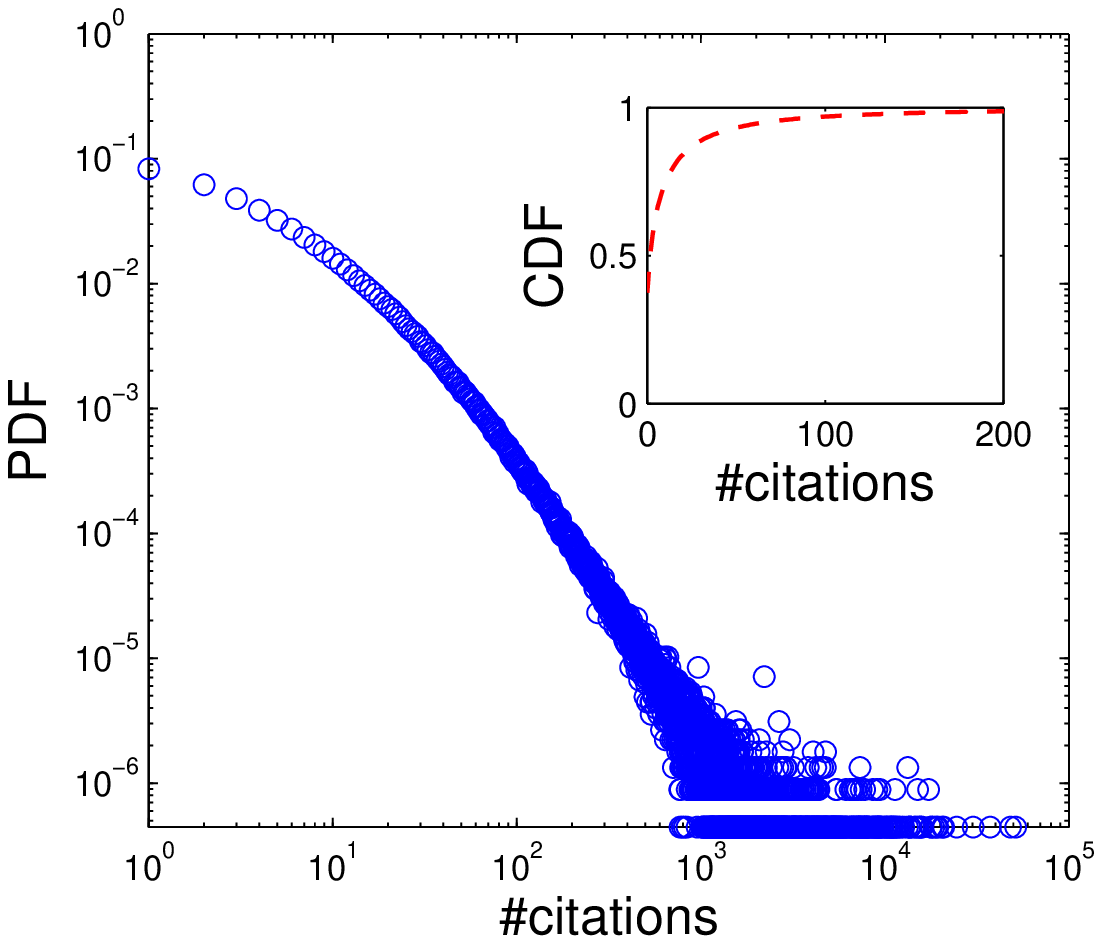}
}
\hspace{-0.1in}
\subfigure[\scriptsize \hindex\ distribution]{
\label{figsub:hindex-powerlaw}
\includegraphics[width=1.75in]{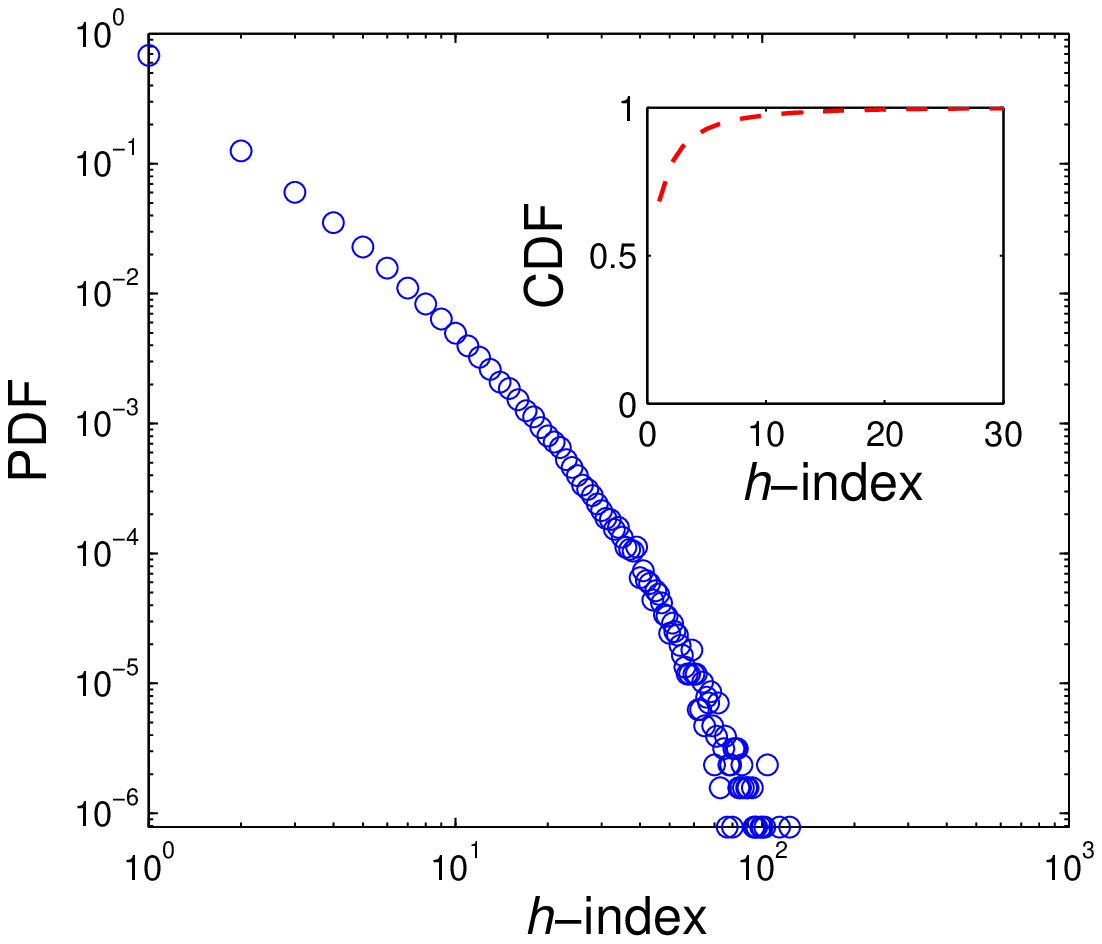}
\hspace{-0.2in}
}
\caption{\label{fig:powerlaw}
{\bf Distributions of the citation counts of papers and the \hindices\ of authors.} 
{
In this dataset, 
7.41\%  
(154,985) of the papers obtain more than 50 citations and 
0.0093\%  
(159) of the researchers have \hindices\ greater than 60.
}
}
\end{figure}

We briefly explore and report the data characteristics of the author-paper-citation data used in this work.
Figure~\ref{fig:powerlaw} shows the distributions of the number of citations for each paper and the \hindex\ of each author.
In our dataset, both metrics follow heavy-tailed distributions (i.e., distributions with a ``tail'' that is ``heavier'' than that of an exponential).
Moreover, only 
7.41\%  
(154,985) of the papers have more than 50 citations, while 
0.0093\%  
(159) of the researchers have an \hindex\ over 60.

A caveat of this work is that by targeting the \hindex, our findings may result in unintended side effects by a principle referred to as Goodhart's Law, 
which essentially warns that ``when a measure becomes a target, it ceases to be a good measure'' \cite{strathern1997improving}.  
Yet, we strongly believe that by deepening the understanding of scientific impact measures, the findings presented in this work can actually help to strengthen the foundations upon which these measures are based, ultimately facilitating their improved use.
\textit{In no way should our research be construed as advocating the use of the \hindex\ or any other measure as a deciding factor in the determination of one's research pursuits}.

\begin{figure}[t]
\centering
\subfigure[\scriptsize \hindex\ vs. \hindex/$\#$papers]{
\label{figsub:paperratio-hratio}
\hspace{-0.2in}
\includegraphics[width=1.75in]{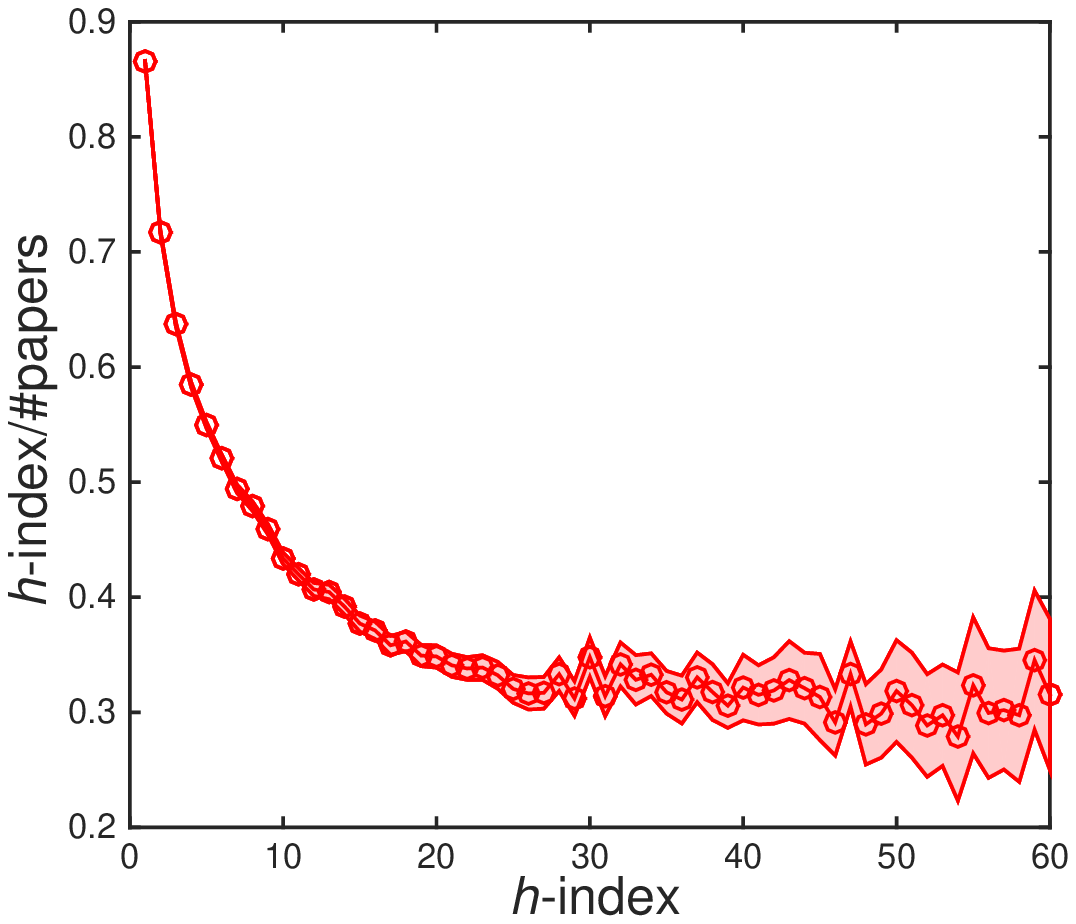}
}
\hspace{-0.1in}
\subfigure[\scriptsize \hindex\ in 2012 vs. 2002\&2007 ]{
\label{figsub:hindex-200x-2012}
\includegraphics[width=1.75in]{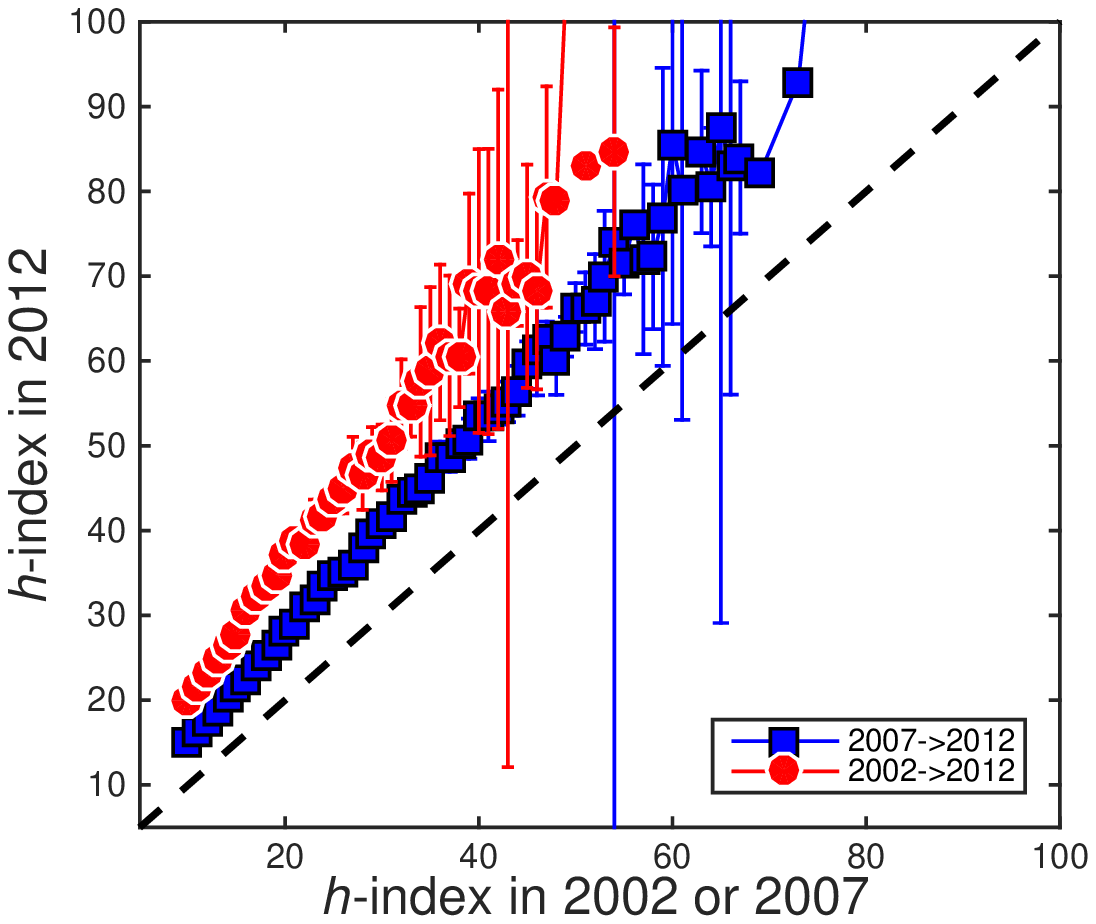}
\hspace{-0.2in}
}
\caption{\label{fig:hindex-stat}
{\bf \hindex\ trends.}
{
(a) The ratio between one's \hindex\ ($\geq$ 20) and her/his number of papers stabilizes at 0.3. (b) The correspondence between one's \hindex\ in 2002 (red line) and 2007 (blue line) and his/her predicted \hindex\ in 2012.
}
}
\vspace{-0.2cm}
\end{figure}

\section{Problem Definition}

Traditionally, the task of scientific impact prediction is formulated as a regression problem for predicting citation counts \cite{Yan:JCDL2012}. 
However the intrinsically heavy-tailed distribution of citation counts, demonstrated in Figure~\ref{figsub:citation-powerlaw}, make such predictions necessarily skewed \cite{Cheng:WWW14,Dong:WSDM2015}. This problem motivates a search for alternate approaches that are more resilient to a skew in citation counts. 
Inspired by the work of \cite{Cheng:WWW14}, which considers the problem of Facebook cascade growth prediction, we formulate the following task:  Given a paper at timestamp $t$, we predict whether that paper will increase its authors' \hindices\ by the future timestamp $t$ + $\Delta t$. 

Realistically, however, the authors' \hindices\ are not static; they may increase during the duration $\Delta t$. 
Figure~\ref{figsub:hindex-200x-2012} shows the comparisons between scholars' \hindices\ in 2002 or 2007 and their corresponding future \hindices\ in 2012. 
In this sense, to solve the scientific impact prediction task above, we need to first infer the future \hindices\ of the paper's authors. 
Thus we formalize two prediction problems, namely future \hindex\ prediction and scientific impact prediction.

\begin{problem}[Future \hindex\ Prediction]
Given the publication corpus $C$ before timestamp $t$ and each author's \hindex\ at $t$, the task is to predict the authors' future \hindices\ at timestamp $t$ + $\Delta t$.
\end{problem}

\begin{definition}[Primary Author]
Given a paper $d \in C$, the primary author of $d$ is defined in two ways: given paper $d$'s author list, take either the author with the highest \hindex\ or the first author on the list. 
\end{definition}

\begin{problem}[Scientific Impact Prediction]
Given the publication corpus $C$ before timestamp $t$, each paper $d \in C$ published by (at or before) $t$, and the primary author's predicted future \hindex, the problem is to predict whether $d$'s number of citations will reach the primary author's future \hindex\ after a given time period $\Delta t$. 
\end{problem}

The major novelty of this approach lies in the formulation of the second problem, i.e., scientific impact prediction, while the first problem serves to facilitate it. 
As formulated, the second problem is composed of two tasks. The first task is to predict for papers published before the current timestamp $t$. For these papers, we have citation counts that have accumulated until $t$. The second task is to predict for those papers published at $t$ without prior information about their citations. Importantly, the problem addresses the above-noted issues with traditional citation count prediction by using a local threshold---the primary author's \hindex---for each paper's future citation count. Figure~\ref{figsub:paperratio-hratio} shows that the ratio between one's \hindex\ ($\geq$ 20) and his or her number of papers stabilizes at about 30\%, allowing us to circumvent the inherent skew of citation counts.

Our proposed problem of scientific impact prediction is fundamentally different from the traditional problem of predicting citation counts \cite{Yan:JCDL2012}. Whereas citation count prediction typically employs regression to predict scientific impact, our problem is to instead predict each paper's future impact conditioned on its authors. 
Though inspired by it, our problem is also entirely different from the cascade growth prediction problem \cite{Cheng:WWW14}, which requires the observation of the first $k$ reshares (here, citations) to predict future reshare counts. 
The chief advantage of our formulation is its general applicability to a variety of real-world tasks, including author \hindex\ and popularity prediction~\cite{Shen:AAAI14}, expert finding and search~\cite{Zhang:07DASFAA,Sun11:VLDB11}, and credit allocation~\cite{Shen:PNAS14,Kleinberg:STOC11}. 
\section{Scientific Impact Factors}
\label{sec:factor}

To quantify scientific impact, it is natural to use the number of citations obtained by each paper and its authors. 
Recall that given a paper $d$, our objective is to predict whether the number of citations $c_{d}$ it obtains within a given time period $\Delta t$ will be larger than its primary author's future \hindex. 
In other words, we aim to model the co-evolution of the primary author's \hindex\ and paper $d$'s citation count over the period $\Delta t$.

\begin{figure}[t]
\centering
\subfigure[\scriptsize \hindex\ vs. $\#$papers]{
\hspace{-0.2in}
\label{figsub:h-numpapers}
\includegraphics[width=1.75in]{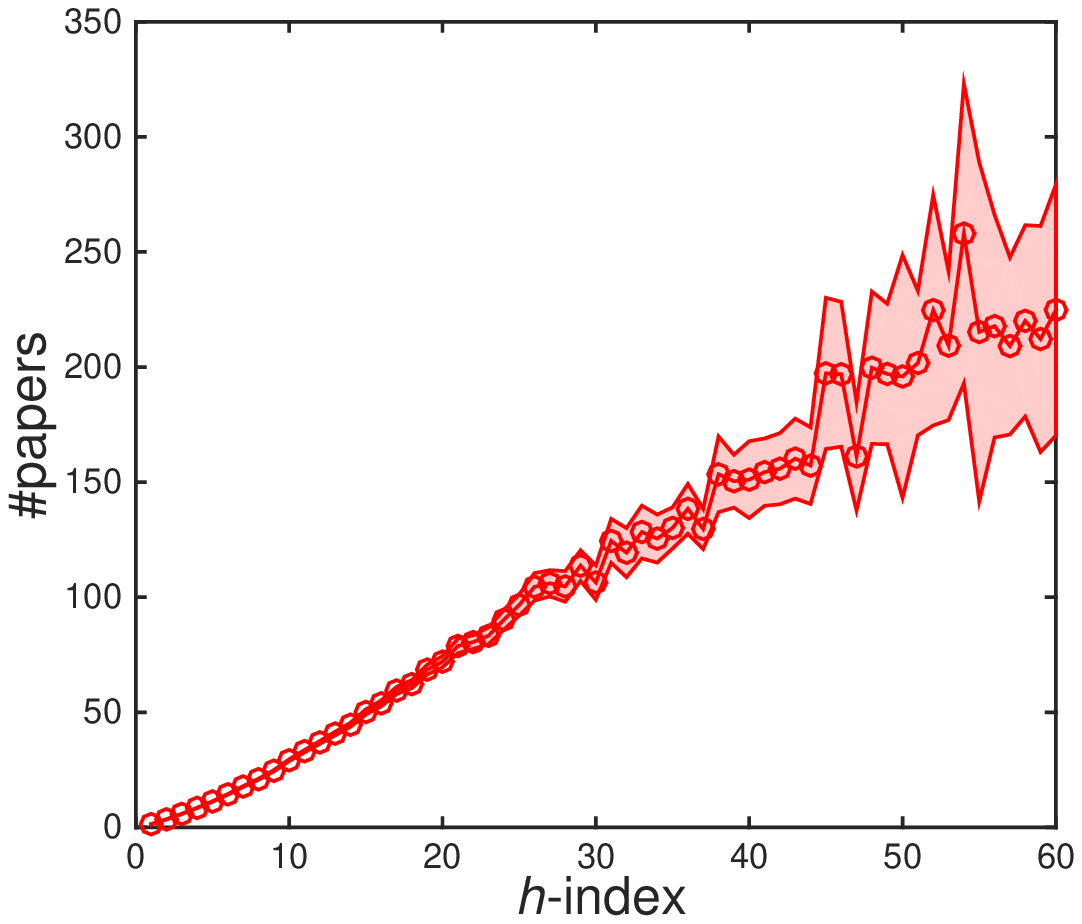}
}
\hspace{-0.1in}
\subfigure[\scriptsize \hindex\ vs. $\#$average citations]{
\label{figsub:h-avec}
\includegraphics[width=1.75in]{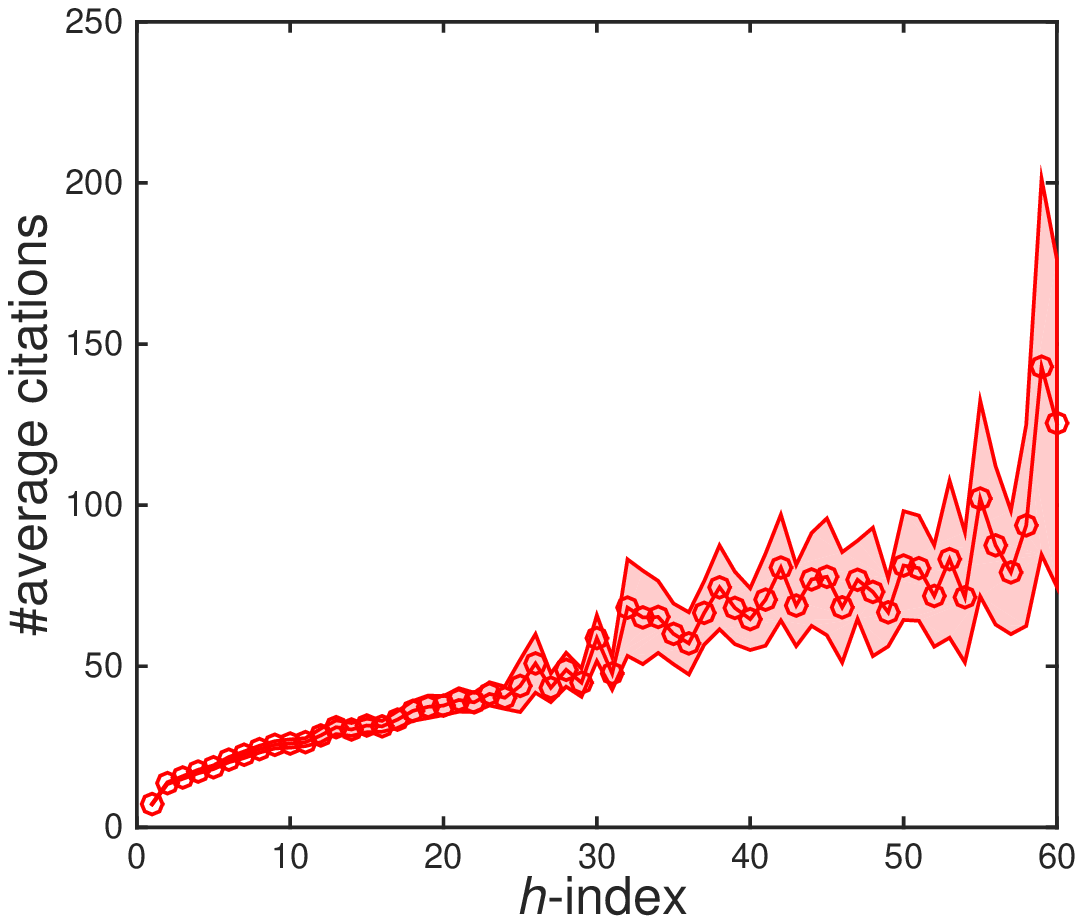}
\hspace{-0.2in}
}
\subfigure[\scriptsize \hindex\ vs. $\#$co-authors]{
\hspace{-0.2in}
\label{figsub:h-numco-authors}
\includegraphics[width=1.75in]{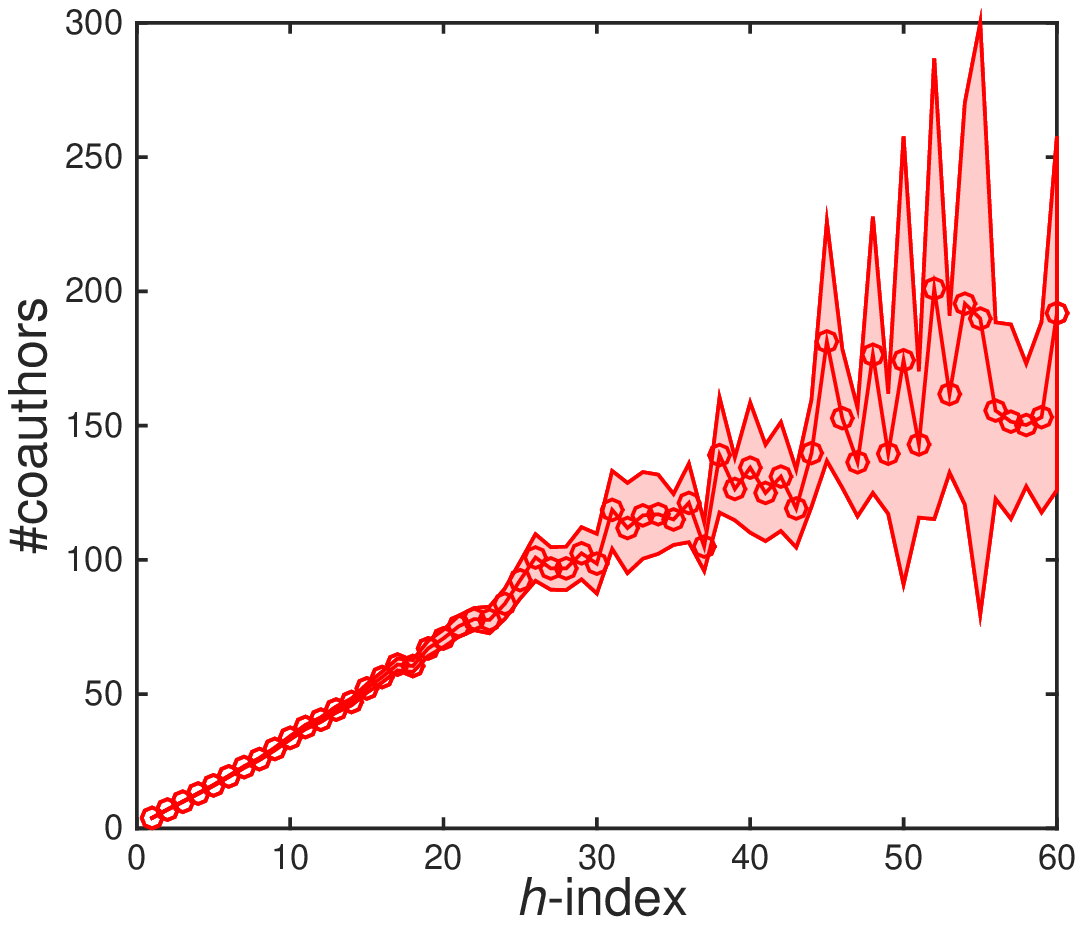}
}
\hspace{-0.1in}
\subfigure[\scriptsize \hindex\ vs. $\#$years]{
\label{figsub:h-years}
\includegraphics[width=1.75in]{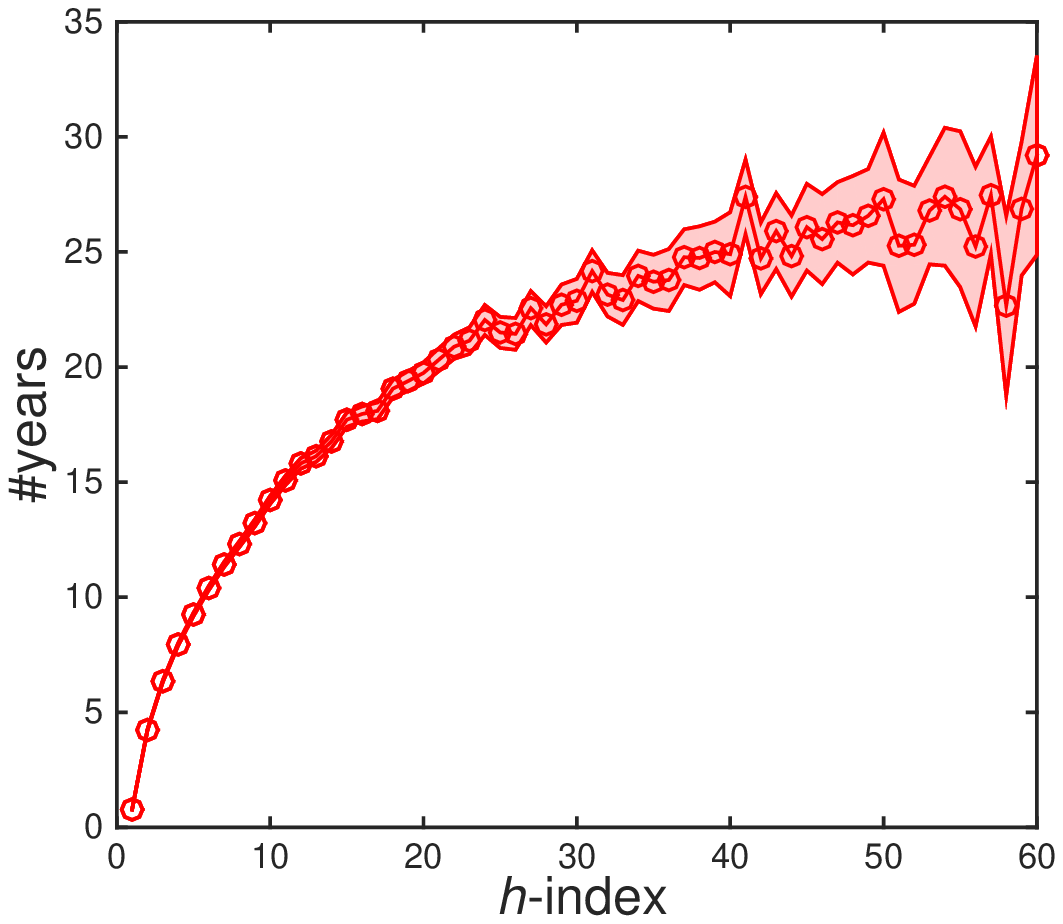}
\hspace{-0.2in}
}
\caption{
\label{fig:citation-h}
{\bf \hindex\ factor correlations. }
{
(a) (c) The numbers of papers and co-authors are highly correlated with a scholar's \hindex. 
(b) The average number of citations for each author is larger than her/his \hindex. 
(d) The rate at which the \hindex\ increases itself increases as the length of time spent in academia becomes longer (\textit{i.e., the rich get richer}). 
Shaded area indicates error bars observed at a 95\% confidence interval. 
}
}
\end{figure}

\begin{table}[t]
\caption{{\bf \hindex\ factor definitions.}
{
Given a researcher's \hindex\ in 2002 and 2007, we study the correlations between several factors and her/his \hindex\ in 2012. 
$cc_{2002}$ and $cc_{2007}$ represent the respective correlation coefficients. 
}
}
\label{tb:hindexfactors}
\centering
\renewcommand\arraystretch{1.35}
\begin{tabular}{l|l|r|r}

\hline

Factor                  & Description                   & $cc_{2002}$ & $cc_{2007}$ \\ \hline
\textit{\hindex}        & Current \hindex               & 0.7838 & 0.9335\\
\textit{num-papers}     & \#papers published            & 0.6518 & 0.7375\\
\textit{num-citations}  & Average \#citations per paper & 0.1486 & 0.2289\\
\textit{num-co}         & \#unique co-authors           & 0.5784 & 0.5992\\
\textit{num-years}      & \#years since first paper     & -0.0855 & 0.1089\\

\hline
\end{tabular}
\vspace{-0.2cm}
\end{table}

\subsection{Factors that Drive One's \hindex\ to Increase}

We first examine the factors that potentially affect the development of scientific scholars' \hindices. 
Acuna et al.~\cite{Acuna:Nature12} and Redner et al.~\cite{Redner:hindex10} have examined the factors that are indicative of the future \hindices\ of small groups of physicists and neuroscientists, respectively. 
As our work focuses on the computer science domain, Table \ref{tb:hindexfactors} provides brief descriptions for five simple factors that we find to have effects on the evolution of computer scientists' \hindices, as well as the correlation coefficients between these factors in 2002 ($\Delta t$=10 years) / 2007 ($\Delta t$=5 years) and the scholars' future \hindices\ in 2012. 

The correlation coefficients provide several observations. 
First, we can observe that researchers' future \hindices\ are highly correlated with their current \hindices, followed by their number of publications and co-authors. 
Second, we notice a potentially counterintuitive phenomenon, wherein the number of citations and years publishing work have surprisingly limited correlations with future \hindices\ vis-\`a-vis other factors. 
Finally, within a shorter timeframe ($cc_{2002}$ vs. $cc_{2007}$), historical and future \hindices\ exhibit high correlations. 

Figure~\ref{fig:citation-h} presents the basic characteristics of scientific impact in terms of \hindex, including counts for an author's number of papers, citations, co-authors, and years conducting research. 
Positive linear relationships are clearly observed between the \hindex\ and the number of papers and co-authors in Figures~\ref{figsub:h-numpapers} and \ref{figsub:h-numco-authors}, respectively. 
Also, Figure~\ref{figsub:h-avec} shows that the average number of citations for each author is larger than his or her \hindex. 
Finally, in Figure~\ref{figsub:h-years}, we examine the interplay between authors' \hindices\ and the length of time they spend in academia (the date difference between one's first and last publications). 
We observe that the increase of \hindex\ is relatively slow upon initially entering academia. 
As one's \hindex\ increases, the accumulations of influence, resources, connections, and publications further drive one's \hindex\ upward, and scientific impact expands at an increasingly rapid rate. 
In other words, the aphorism that ``the rich get richer'' is readily observed in academia, whereby the influence of individuals who have already accumulated a great deal of influence increases at a disproportionally quick rate. 
All characteristics are observed at a 95\% confidence interval.

\begin{table*}[t]
\caption{{\bf Factor definitions and correlations.}
{
We employ six categories of 24  factors, composed of author, topic, reference, social, venue, and temporal attributes. 
$cc$ denotes the correlation coefficients. 
max-\hindex\ denotes the highest or ``maximum'' \hindex\ among the authors' \hindices. 
$P_{new}^{max}$ denotes the case where we define the max-\hindex\ author as the primary author of a newly published paper. 
$P_{new}^{first}$ denotes the case where we define the first author as the primary author. 
}
}
\label{tb:factors}
\centering
\scriptsize
\renewcommand\arraystretch{1.35}
\begin{tabular}{@{}l|l|l|r|r|r|r}

\hline

   \multirow{2}{*}{}    
  &\multirow{2}{*}{Factor} 
  &\multirow{2}{*}{Description} 
  &\multicolumn{2}{c|}{$P_{new}^{max}$}
  &\multicolumn{2}{|c}{$P_{new}^{first}$}\\
  \cline{4-7} &&& $cc_{2002}$ & $cc_{2007}$ & $cc_{2002}$ & $cc_{2007}$ \\ \hline

\multirow{7}*{Author} & \textit{A-first-max} & The first author's \hindex.   & 0.0309 & 0.0728 & 0.1102 & 0.1998\\ 
& \textit{A-ave-max}     & The average \hindex\ of all authors.              & 0.0435 & 0.0999 & 0.0670 & 0.0264 \\ 
&\textit{A-sum-max}      & The sum of \hindices\ of all authors.             & 0.1589 & 0.1585 & 0.1801 & 0.1915 \\ 
& \textit{A-first-ratio} & The ratio between max-\hindex\ and \#papers attributed to the first author.    & 0.0161 & -0.0365 & 0.2904 & 0.3232 \\ 
&\textit{A-max-ratio}    & The ratio between max-\hindex\ and \#papers attributed to the primary author.  & 0.2866 & 0.2423 & 0.2601 & 0.2285 \\ 
&\textit{A-num-authors}  & The number of authors of the given paper.                                      & 0.0878 & 0.0617 & 0.1359 & 0.0668\\ 

\hline

\multirow{7}*{Content} & \textit{C-popularity} & The \#average-citations over different topics (see Eq.~\ref{eq:popularity}).      & 0.2085 & 0.0741 & 0.2590 & 0.0628 \\ 
&\textit{C-novelty}          & The topic novelty of this paper (see Eq.~\ref{eq:novelty}).                                         & 0.1192 & 0.0807 &0.1262 & 0.0763\\ 
&\textit{C-diversity}        & The topic diversity of this paper (see Eq.~\ref{eq:diversity}).                                     & 0.1852 & 0.0712 & 0.2498 & 0.0716 \\ 
&\textit{C-authority-first}  & The consistence between the first author's authority and this paper (see Eq.~\ref{eq:authority}).   & 0.3537 & 0.4346 & 0.3408 & 0.3490 \\ 
&\textit{C-authority-max}    & The consistence between the primary author's authority and this paper.                              & 0.3265 & 0.3874 & 0.3420 &  0.3667\\ 
&\textit{C-authority-ave}    & The average consistence between each author's authority and this paper.                             & 0.3611 & 0.4359 & 0.3623 & 0.3865 \\ 

\hline

\multirow{2}*{Venue} & \textit{V-\hindex} & The venue's \hindex.                              & 0.2557 & 0.2940 &0.2400 & 0.2351 \\  
&\textit{V-citation} & The \#average-citations of papers published in this venue.               & 0.3357 & 0.3506 & 0.3058 & 0.3194 \\  

\hline

\multirow{4}*{Social}  & \textit{S-degree} & The number of co-authors of the paper's authors.                           & 0.0314  & -0.0393 & 0.0340 & 0.0454\\ 
&\textit{S-pagerank}   & The PageRank values of the paper's authors in the weighted collaboration network.              & -0.0341 & -0.0782 & 0.0500 & 0.1317\\ 
&\textit{S-h-coauthor} & The average \hindex\ of co-authors of the paper's authors.                                    & 0.0750  & 0.0976 & 0.0148 & 0.0206\\ 
&\textit{S-h-weight}   & The weighted average \hindex\ of co-authors of the paper's authors. & 0.0639  & 0.0861 & 0.0006 & 0.0166\\  

\hline

\multirow{2}*{Reference} & \textit{R-\hindex} & The references' \hindex.    & 0.1405 & 0.1562  & 0.1204 &0.1103\\  
&\textit{R-citation} & The \#average-citations.                               & 0.0858 & 0.0420 & 0.0635 & 0.0150\\ 

\hline

\multirow{4}*{Temporal} & \textit{T-ave-h} & The average $\Delta$\hindices\ of the authors between now and three years ago.   & 0.2528  & 0.2616 & 0.1740 & 0.1819\\  
&\textit{T-max-h}   & The maximum $\Delta$\hindex\ between now and three years ago.                                           & 0.2539  & 0.2027 & 0.2426 & 0.2032\\ 
&\textit{T-h-first} & The $\Delta$\hindex\ of the first author between now and three years ago.                               & 0.2109  & 0.2188 & 0.1737 & 0.0907\\ 
&\textit{T-h-max}   & The $\Delta$\hindex\ of the max-\hindex\ author between now and three years ago.                        & 0.2117  & 0.1504  & 0.2012 & 0.1603\\ 

\hline

\end{tabular}
\end{table*}

\begin{figure*}[t]
\centering
\subfigure[\scriptsize Author factors]{
\label{figsub:cc-h-author}
\includegraphics[width=2.25in]{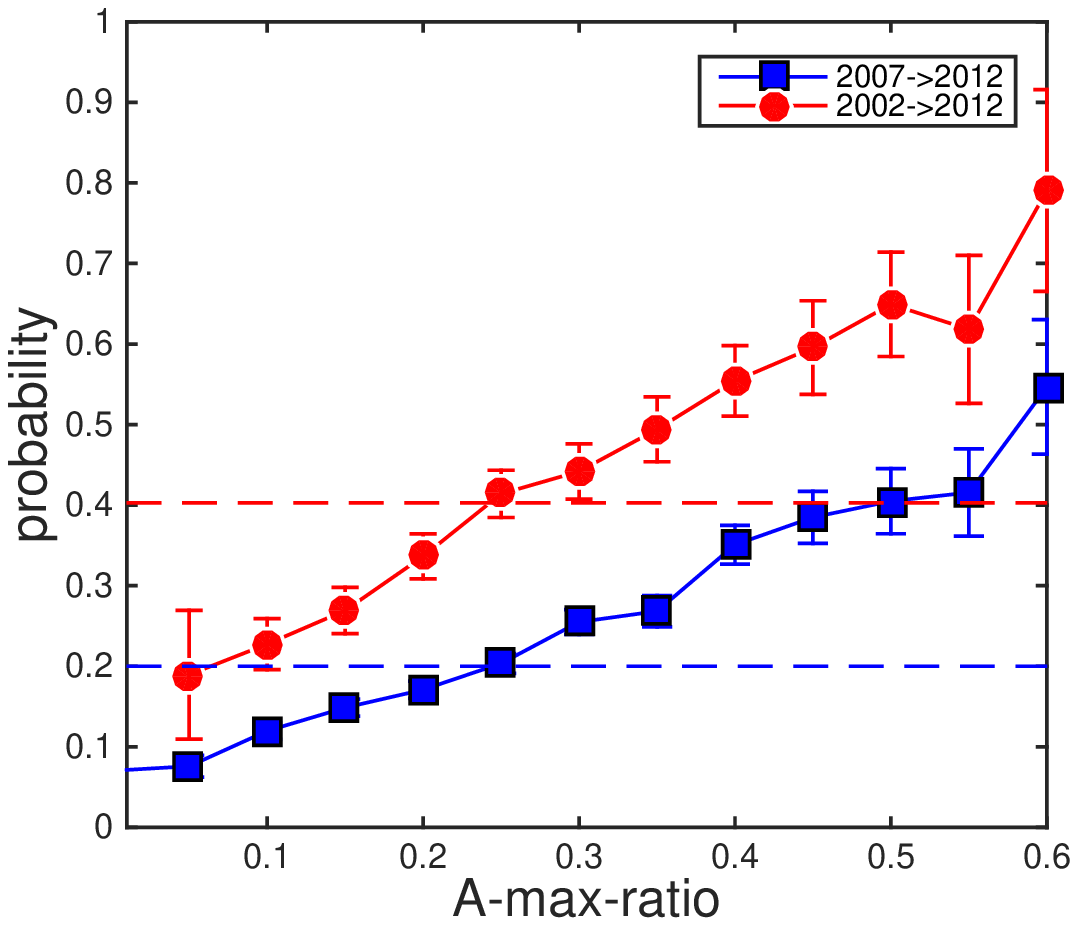}
}
\subfigure[\scriptsize Content factors]{
\label{figsub:cc-h-content}
\includegraphics[width=2.25in]{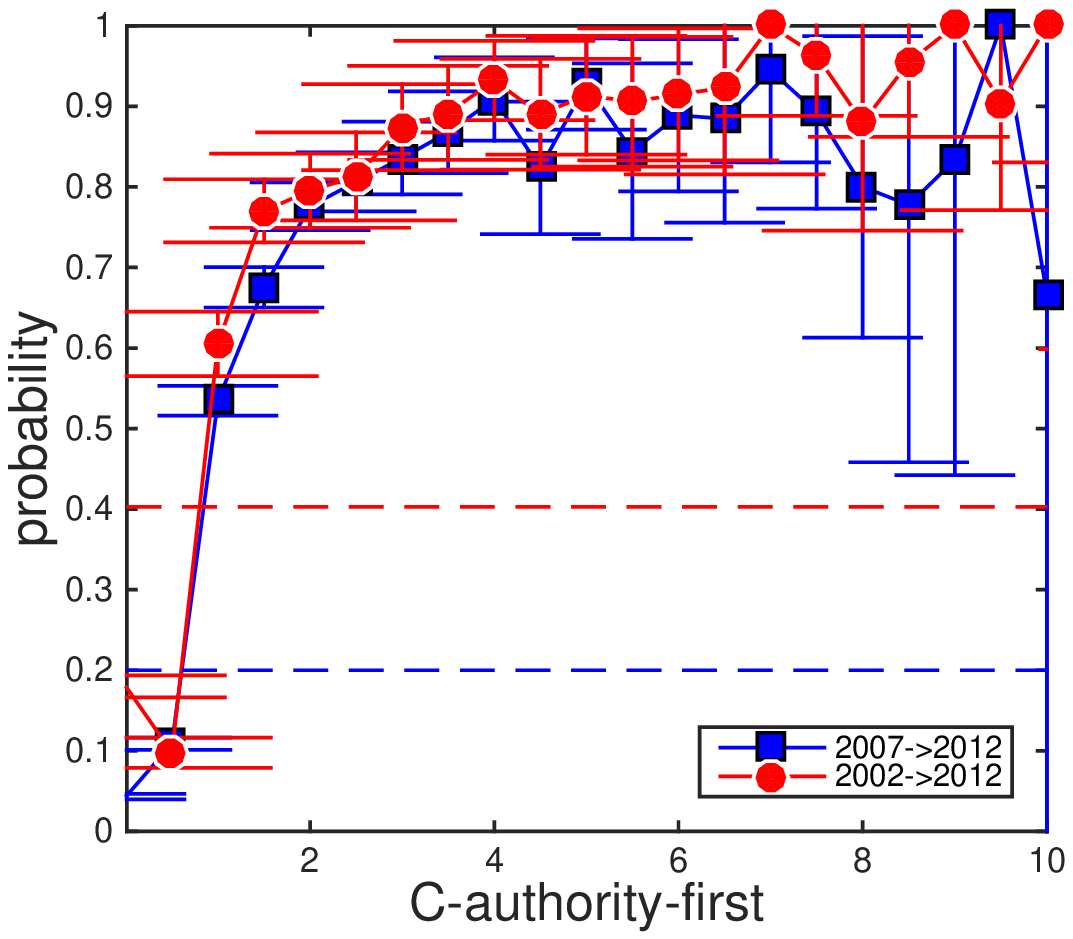}
}
\subfigure[\scriptsize Venue factors]{
\label{figsub:cc-h-venue}
\includegraphics[width=2.25in]{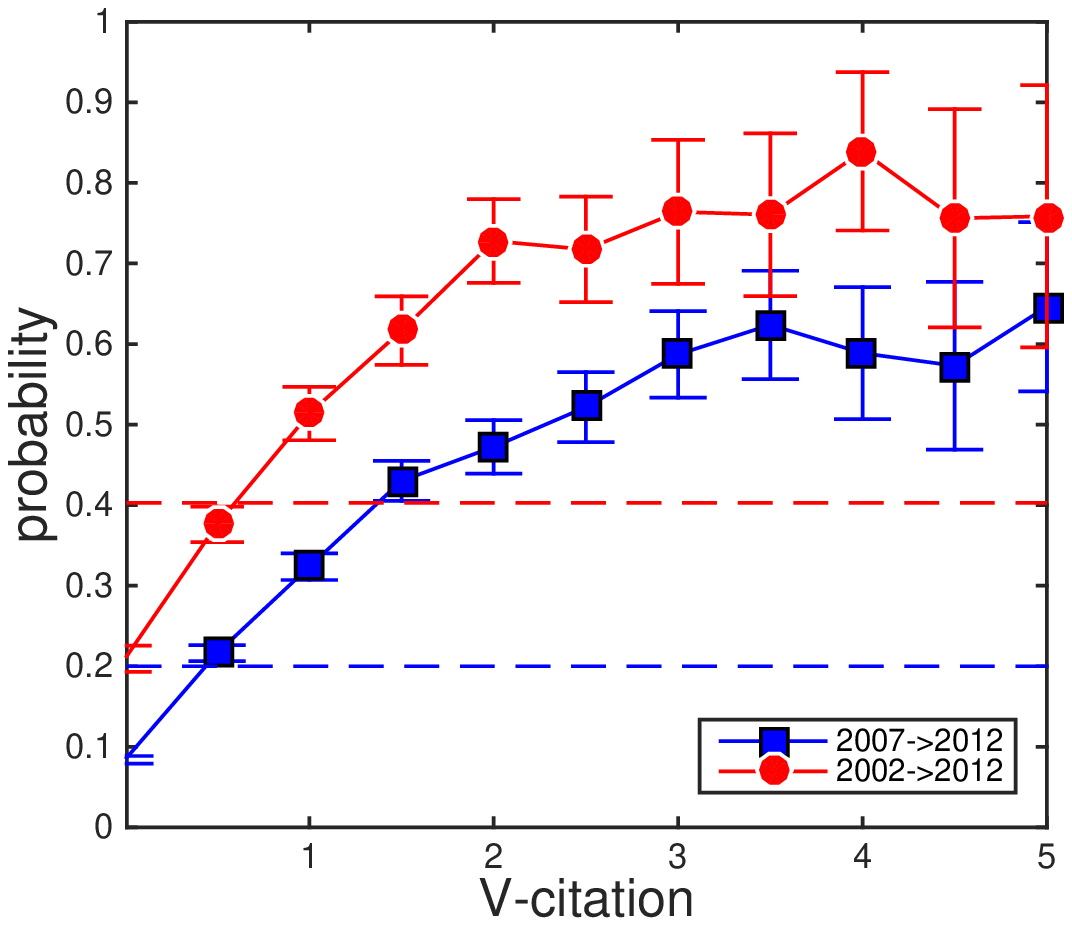}
}
\subfigure[\scriptsize Social factors]{
\label{figsub:cc-h-social}
\includegraphics[width=2.25in]{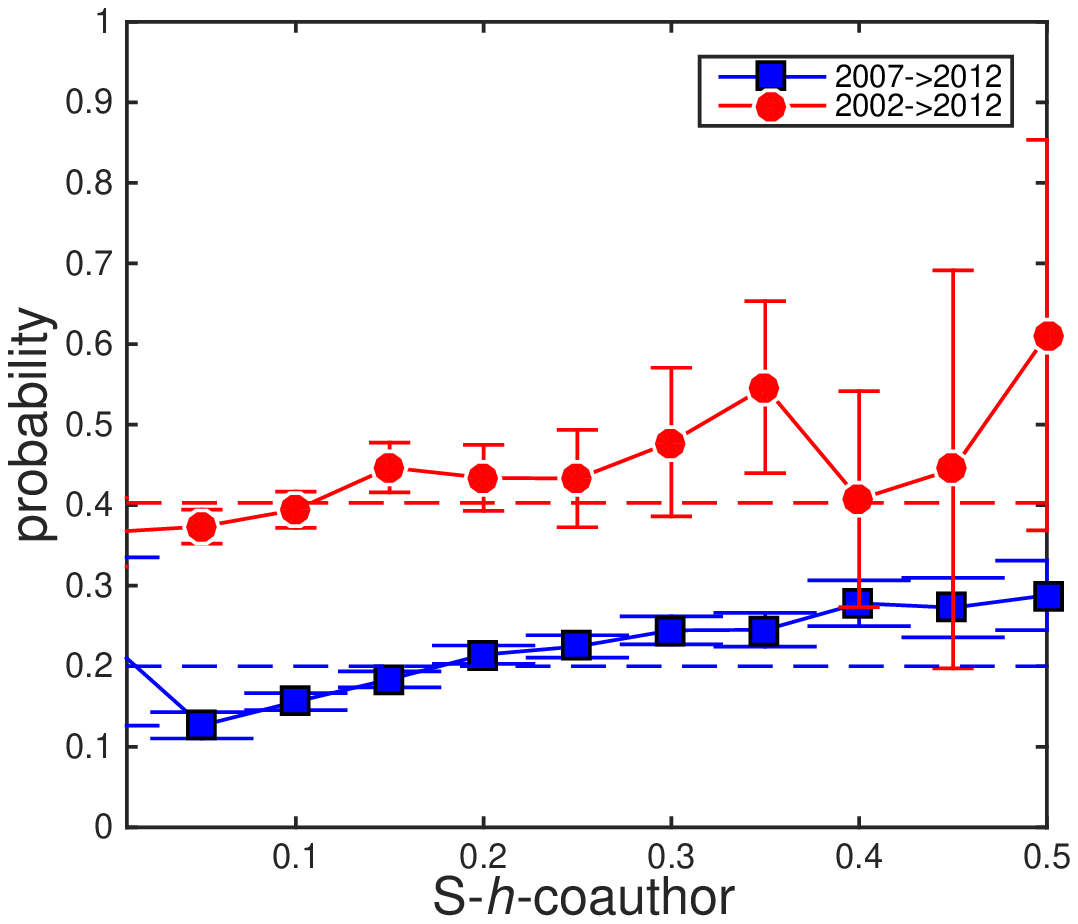}
}
\subfigure[\scriptsize Reference factors]{
\label{figsub:cc-h-ref}
\includegraphics[width=2.25in]{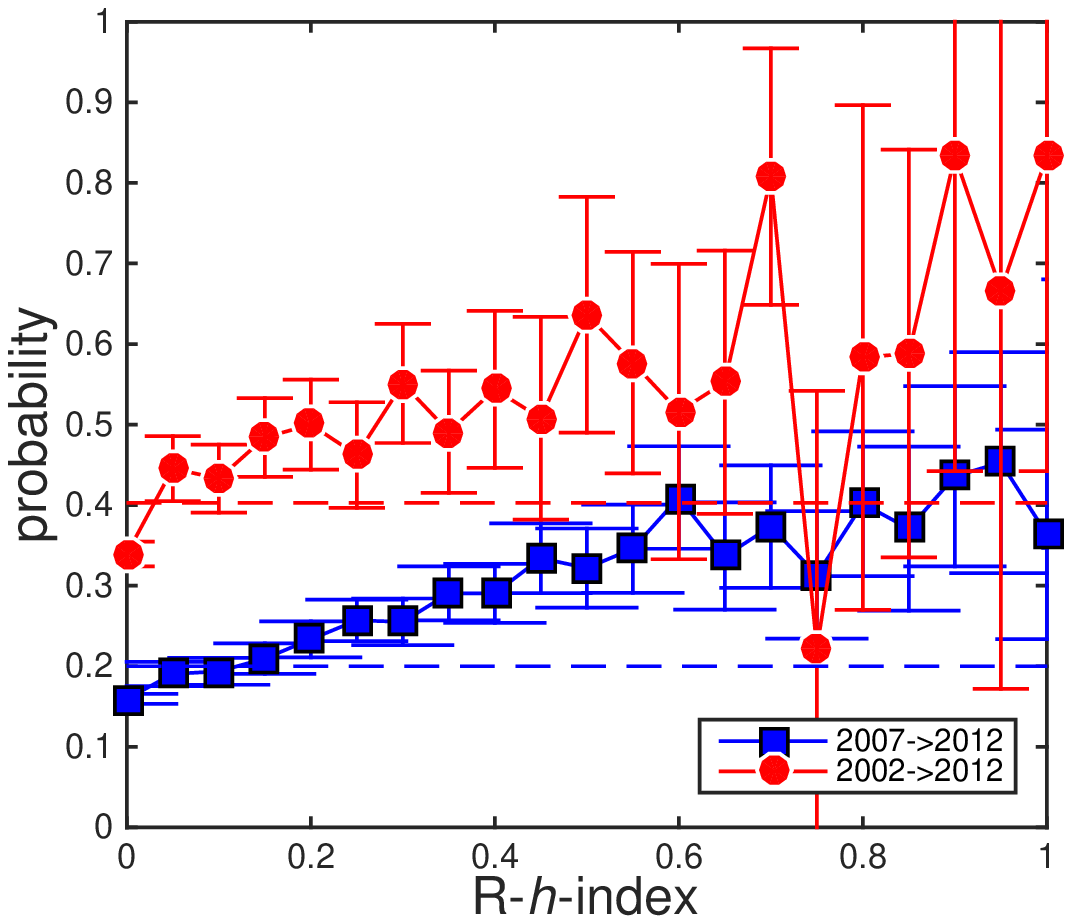}
}
\subfigure[\scriptsize Temporal factors]{
\label{figsub:cc-h-temp}
\includegraphics[width=2.25in]{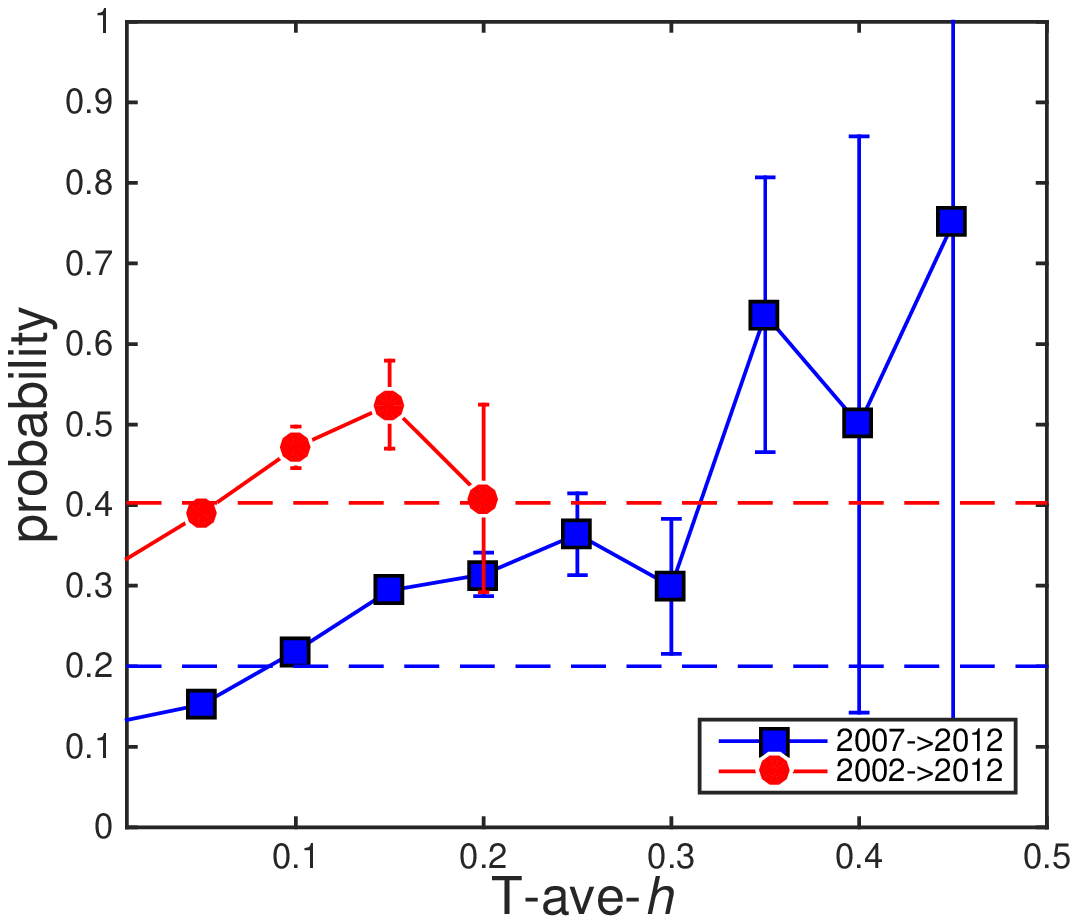}
}
\caption{\label{fig:cc-h}
{\bf Factor response curves with $\Delta t = 5$ or $10$ for $P_{new}^{max}$.} 
\scriptsize{
$x$-axis: factor value; 
$y$-axis: probability that a paper published at time $t$ will increase its primary author's \hindex\ by 2012. 
All response probabilities are observed at a 95\% confidence interval. 
}
}
\vspace{-0.25cm}
\end{figure*}

\subsection{Factors that Drive Papers to Increase \hindex}

We further investigate the factors that drive a paper's citation count to exceed its primary author's \hindex, including the paper's author(s), content, publication venue, and references, as well as social and temporal effects related to its author(s).
Table \ref{tb:factors} lists the six diverse groups of factors investigated in this work, as well as the correlation coefficients between the factors of papers published in 2002 ($\Delta t=10$) / 2007 ($\Delta t=5$) and whether their citation counts are greater than or equal to the primary authors' \hindices\ in 2012.  
Figure \ref{fig:cc-h} shows the response curve of the most important factor for each group of factors (as evaluated by correlation coefficients in Table \ref{tb:factors}) when considering the max-\hindex\ author as the primary author.

\vpara{Author Factors.}
The prediction task for each paper naturally depends on the authors themselves, including both the primary author and his or her co-authors. 
Prior work has been devoted to examining the interplay between scientific impact (number of citations) and the average values of authors' attributes \cite{Yan:JCDL2012,Castillo:2007}. 
Given our problem formulation, in addition to these factors, for each paper we investigate the attributes of the primary author (e.g., the ratio of the author's previous papers that contribute to his/her \hindex). 
Additionally, as the first author of a publication usually leads the collaboration and may have considerable influence on its scientific impact, we consider the probability that the number of citations obtained by each of the first author's previous publications is greater than the primary author's \hindex. 
Of course, as a paper is the sum of all authors' contributions, the combined impact of all co-authors may influence a paper's quality and popularity. 
Thus the sum of all authors' \hindices\ is used to simulate their overall impact. 
Due to self-citation behavior, the author's productivity (i.e., the number of her/his previous publications) also has a positive effect on the paper's future citation counts~\cite{Bethard:CIKM2010}.

\vpara{Content Factors.}
Aside from the attributes of its authors, another intuitive factor affecting a paper's success is its content. 
Topic modeling is a widely used method for extracting and mining the content of literature and can be used to extract ``topics'' that occur in a collection of documents. One of the most popular topic modeling methods is known as Latent Dirichlet Allocation (LDA), a generative probabilistic approach that views each document as a mixture of various topics \cite{Blei:03}.
Similar to previous work on modeling citation counts~\cite{Yan:JCDL2012}, we run a 100-topic LDA model on the title and abstract of the corpus $C$ published before time $t$ and the target papers published at time $t$, which returns the probability distribution $p(z|d)$ over topics $z\in Z$ assigned for each paper $d$.
We denote a target paper $d$ at time $t$ as $d_t$, and we define several features based on each paper's topic distribution, including popularity, novelty, diversity, and authority. 
We provide details on these features next.

First, we consider that as popular topics tend to attract more attention and resources than relatively unpopular ones, it is relatively easy for papers related to such topics to accrue citations. 
To capture this effect, we quantify the popularity of each topic $z$ across the overall corpus by
$\scriptsize popularity(z) = \sum_{d \in C} p(z|d) \times c_d,$ 
where $p(z|d)$ is the probability that paper $d$ distributes on topic $z$ and $c_d$ is the number of citations $d$ collects until the timestamp $t$.
The popularity of a target paper $d_t$ (paper $d$ at time $t$) is then defined as:
\beq{
\small
\textit{C-popularity}(d_t) = \sum_{z \in Z} popularity(z) \times p(z|d_t).
\label{eq:popularity}
}

Second, a paper's novelty is an essential factor when assessing its contribution to the scientific community. 
Previous work assumes that the novelty of an article can be determined by measuring the difference between its content and that of its references~\cite{Yan:JCDL2012}.
We utilize the Kullback-Leibler divergence~\cite{Kullback:51} to capture the sum of the difference between $d_t$'s topic distribution and the topic distribution of each of its references. Specifically, we define the novelty of paper $d_t$ as
\beq{
\small
\textit{C-novelty}(d_t) = \frac{\sum_{d_r \in R} KL(p(Z|d_t), p(Z|d_r)) }{|R|}, \label{eq:novelty}
}

\noindent where $KL(p(Z|d_t), p(Z|d_r)) = \sum_{z\in Z} \log \frac{p(z|d_t)}{p(z|d_r)} p(z|d_t)$ and $R$ is the set of $d_t$'s references.

Third, the topic diversity of a paper, defined as the breadth of its topic distribution, is able to distinguish between different types of papers, such as surveys and technical work.
We follow the definition of diversity in~\cite{Yan:JCDL2012} as calculated by Shannon entropy:
\beq{
\small
\textit{C-diversity}(d_t) = \sum_{z\in Z} -p(z|d_t)\log p(z|d_t).
\label{eq:diversity}
}

Fourth, Kleinberg has pointed out that in a hyperlinked web environment, a ``good'' authority represents a page that is linked to by many hubs~\cite{Kleinberg:99}. 
Similarly, academic authority can be designated by being highly cited by many other researchers in a specific domain of expertise.
To measure the authority of researcher $a$ on topic $z$, we propose the following definition: 
$authority(a,z) = \sum_{d \in C_a} p(z|d) \times c_d,$ 
where $C_a$ is the researcher $a$'s previous publications. 
Therefore, given the target paper $d_t$, the author's authority is distributed over the topic distribution of $d_t$. Formally,
\beq{
\small
\textit{C-authority}(d_t, a) = \sum_{z\in Z} p(z|d_t) \times authority(a, z).
\label{eq:authority}
}

This definition of authority follows from the intuition that a correspondence between a paper's topic distribution and its authors' expertise can help ensure its quality.

\vpara{Venue Factors.}
Top venues attract high-quality submissions, and high-quality submissions elevate the reputation of their respective venues. 
Google Scholar metrics show that different venues have large differences in their \textit{h5}-indices\footnote{\scriptsize \url{http://scholar.google.com/citations?view_op=top_venues} Accessed on Nov. 25th, 2014.} (the \hindex\ computed only from articles published within the last 5 complete years), a measure of venue impact.
For example, in the field of data mining and analysis, the top three venues are ACM SIGKDD, IEEE TKDE, and ACM WSDM, with \textit{h5}-indices of 69, 57, and 54, respectively. By contrast, most other venues in this field typically have \textit{h5}-indices between 10 and 20.
In light of these differences, we engage in the investigation of how different venues influence the probability that a paper contributes to its author's \hindex.
Two heuristic metrics are examined, namely (1) the average number of citations each paper in the venue collects and (2) the ratio between the number of papers in the venue with at least max-\hindex\ citations to the venue's total number of papers.
Every researcher aims to publish scientific results in well-respected journals and conferences, so our intuition is that top venues help researchers spread their scientific impact and, more specifically, to increase the citation counts of their papers, which further offers a potential to increase their \hindices.

\vpara{Social Factors.}
Previous studies have shown that researchers display a tendency to cite their co-authors' work \cite{Bethard:CIKM2010}.
As shown in Figure~\ref{figsub:h-numco-authors}, our investigations reveal that a researcher's \hindex\ is also positively correlated with his or her total number of collaborators/co-authors.
To explore this trend, we extract a weighted collaboration network from the dataset, where each author is denoted as a node and each link between two nodes is connected if the researchers have collaborated with each other.
The weight of each link is defined as the frequency of collaboration.
We then extract four features for each node (author) from the collaboration network, including the number of co-authors (degree), the PageRank score, the average \hindex\ of co-authors, and the weighted average \hindex\ of co-authors.
For a given paper, the highest values among its authors for these four metrics are used as social factors.

\vpara{Reference Factors.}
The scientific impact of a scholarly work is often quantified by its respective citation count. 
The more times a publication is cited by others, the greater its assumed impact. 
Conversely, as most scientific research is undertaken by ``standing on the shoulder of giants,'' we ask whether highly cited papers actually tend to acknowledge the previous ``giants'' upon whom they stand.
Two intuitive factors are used to evaluate this question, namely (1) the ratio of a paper's references that have at least max-\hindex\ citations to the paper's total number of references and (2) the average number of citations accumulated by the paper's references.

\vpara{Temporal Factors.}
Just as fast-rising phenomena typically attract the attention of crowds more easily, a ``rising star'' in academia can attract wide publicity. Previous work has found that temporal information can be a powerful factor in modeling scientific impact~\cite{Bethard:CIKM2010,Yan:JCDL2012}, so it is straightforward to assume that the speed at which an author's \hindex\ grows should affect the rate at which the author's papers contribute to his or her \hindex.
To capture this effect, we examine the increase of authors' \hindices\ within the past three years.
Specifically, we consider four temporal factors, including the \hindex\ changes of the first author, the max-\hindex\ author, and the average change and maximum change among all authors. 
The specific definitions are shown in Table \ref{tb:factors}.

\subsection{Existing Factors for Previous Papers}

Besides the above-examined factors, 
which generally drive papers to increase authors' \hindices, 
we discuss several factors that are extracted from the existing citation information for papers published before time $t$. 
For each paper, we consider three intuitive factors: (1) the total number of citations the paper has accrued until $t$; (2) the average number of citations the paper has accrued per year until $t$; and (3) the length of time between the paper's publication date and $t$.

The correlation of each factor with the target variable is provided in Table \ref{tb:existingfactors}. We observe that, from among these factors, the average number of citations per year that each paper has accrued before $t$ is most highly correlated with the probability that the paper will increase its primary author's future \hindex\ at time $t$ + $\Delta t$.

\subsection{Summary}
Drawn from the correlation analysis above, we provide the following intuitions relating to academia: 

First, a research scholar's future \hindex\ is highly correlated with his or her current impact---namely, the researcher's \hindex---rather than the number of citations each of his or her publications collect or the length of his or her academic career. 

Second, 
a scientific researcher's authority on a topic is the most decisive factor in facilitating an increase in his or her \hindex. 
This coincides with the fact that the society fellows (e.g., NAS/NAE membership) or lifetime honors (e.g., Turing Award) are typically conferred for contributions to a particular topic or domain.

Third, the 
reputation 
of the venue in which a given paper is published is another crucial factor in determining the probability that it will contribute to its authors' \hindices.
Top venues distinguish one's work as outstanding and expand one's scientific impact; gradually, one's impact can further help to increase the venue's prestige.

Finally, while people 
in social society often follow vogue trends, publishing on an academically ``hot'' but unfamiliar topic is unlikely to further one's scientific impact, at least as measured by one's \hindex.

\begin{table}[t]
\caption{{\bf Existing factor definitions and correlations.}
\scriptsize{
Factors extracted from existing citation information for papers published before time $t$ (where $t$=2002/2007). 
}
}
\label{tb:existingfactors}
\centering
\scriptsize
\renewcommand\arraystretch{1.35}
\begin{tabular}{l|l|r|r|r|r}

\hline
   \multirow{2}{*}{Factor} 
  &\multirow{2}{*}{Description} 
  &\multicolumn{2}{c|}{$P_{new}^{max}$}
  &\multicolumn{2}{|c}{$P_{new}^{first}$}\\
  \cline{3-6} && $cc_{2002}$ & $cc_{2007}$ & $cc_{2002}$ & $cc_{2007}$ \\ \hline

\textit{E-numc}        & \#citations                    & 0.1656 & 0.2352 & 0.1509 & 0.2029\\
\textit{E-numc-ave}    & \#ave-c per year   & 0.1913 & 0.3203 & 0.1579 & 0.2600\\
\textit{E-num-years}   & \#publication-years & 0.0140 & 0.0856 & 0.0103 & 0.0415\\

\hline
\end{tabular}
\vspace{-0.2cm}
\end{table}

\section{Scientific Impact Prediction}
\label{sec:exp}

In this section, we demonstrate the predictability of scientific impact in two parts. 
First, we predict the future \hindices\ of scientific scholars. 
Second, given the estimated future \hindices, we determine whether a previously (\textit{$P_{old}$}) or newly (\textit{$P_{new}$}) published paper will contribute to its primary author's \hindex\ within a given timeframe. 
 
\subsection{Experimental Setup}
Our primary task is to predict whether a paper published by (at or before) timestamp $t$ will contribute to the future \hindex\ of its primary author within a given time period $\Delta t$. 
To accomplish this, we need to first estimate the author's \hindex\ at $t$ + $\Delta t$ based on data observed at time $t$. 
For example, by setting $t=2007$, $\Delta t=5$ years, and the minimum \hindex\ of the primary author to 10, we collect one set of papers (\textit{$P_{new}$}) published in 2007 and another set of papers published before 2007 (\textit{$P_{old}$}). We then extract the features from the corpus observed at 2007 and observe whether the number of citations for each paper in these two sets is larger than or equal to the future \hindex\ of its primary author in 2012 (the last year represented in our dataset).

\subsection{Predicting Future \hindices}

\vpara{Methods.}
Similar to the previous work of~\cite{Acuna:Nature12}, wherein Acuna et al. propose a method to infer the future \hindices\ of neuroscientists, our \hindex\ prediction problem is formulated as a regression task. For this task, we use linear regression, primarily due to its effectiveness, simplicity, and interpretability. 
The features used here contain the factors detailed in Table~\ref{tb:hindexfactors}. 
To quantitatively evaluate the model predictions, we report the performance in terms of the coefficient of determination ($R^2$)~\cite{R2:Magee90} and the mean absolute error (MAE).

\begin{figure}[t]
\centering
\subfigure[\scriptsize Predictive performance ($R^2$)]{
\hspace{-0.2in}
\label{figsub:pred-r2}
\includegraphics[width=1.75in]{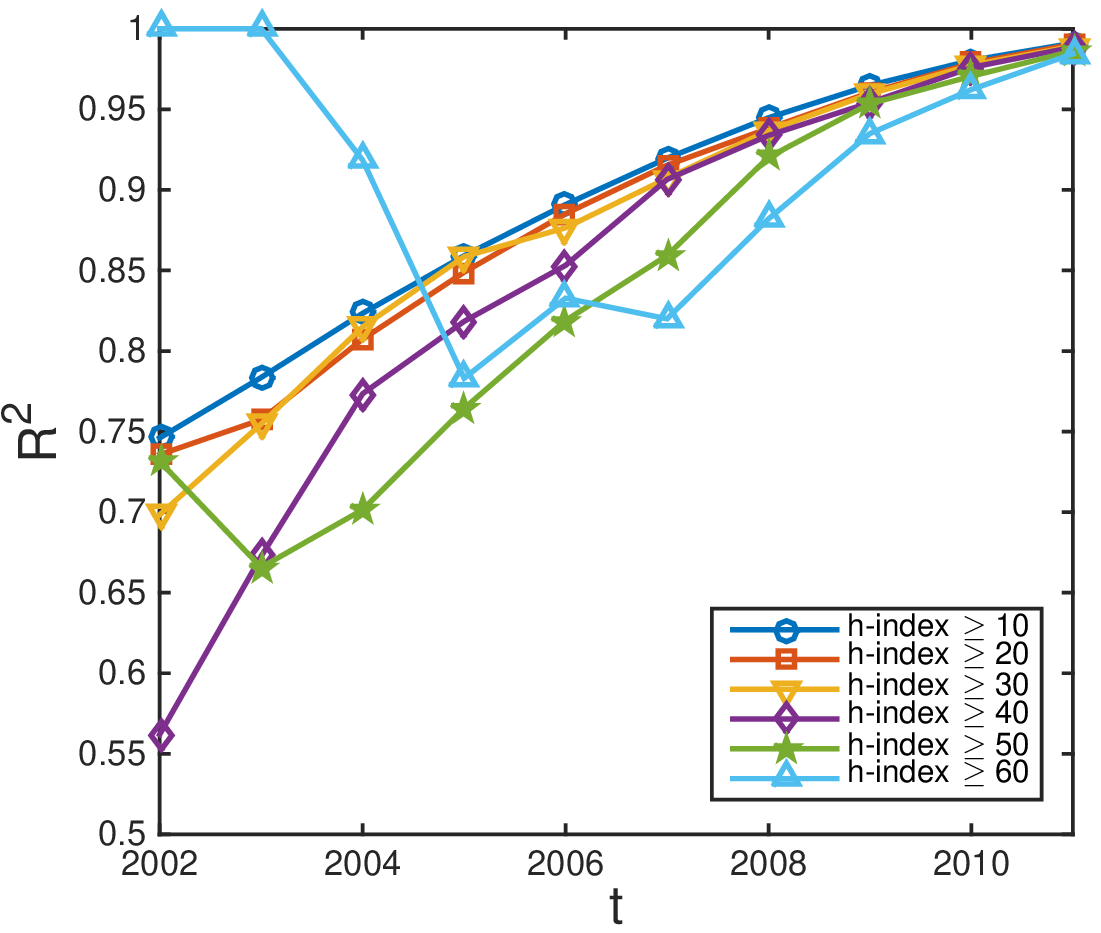}
}
\hspace{-0.1in}
\subfigure[\scriptsize Predictive performance (MAE)]{
\label{figsub:pred-mae}
\includegraphics[width=1.75in]{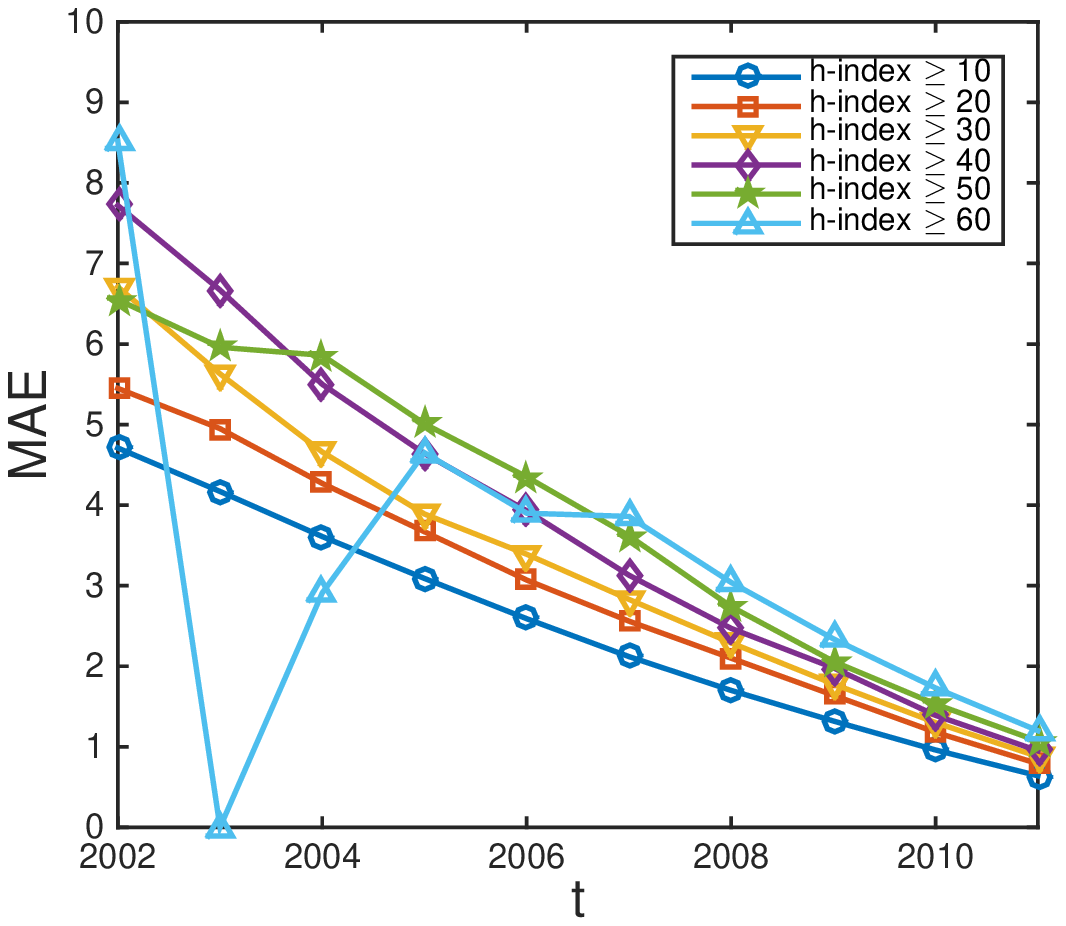}
\hspace{-0.2in}
}
\caption{\label{fig:varying-para-r2}
{\bf Performance for predicting future \hindices.} 
}
\vspace{-0.3cm}
\end{figure}

\vpara{Prediction Results.} 
We present the extent to which research scholars' future \hindices\ can be inferred from their previous publication records. 
Figure~\ref{fig:varying-para-r2} reports the predictive performance in terms of $R^2$ and MAE. 
On the one hand, the rising lines in Figure \ref{figsub:pred-r2} and the descending lines in Figure \ref{figsub:pred-mae} as $t$ increases both imply that our prediction task is easier when given a shorter timeframe.
That is, future \hindices\ are more predictable 
when the future is close to $t$. 
Our observations agree with the intuition that the variability in the development of researchers' \hindices\ increases with a large prediction timeframe. 
On the other hand, the figure generally suggests that our prediction task is more difficult for authors with high \hindices. 
Intuitively, as an author's \hindex\ increases, the variability in the development of his or her scientific impact also increases, which results in an increasingly challenging prediction task.

\begin{figure}[t]
\centering

\subfigure[\scriptsize Prediction from 2002 to 2012]{
\label{figsub:pred-h-2002}
\hspace{-0.2in}
\includegraphics[width=1.75in]{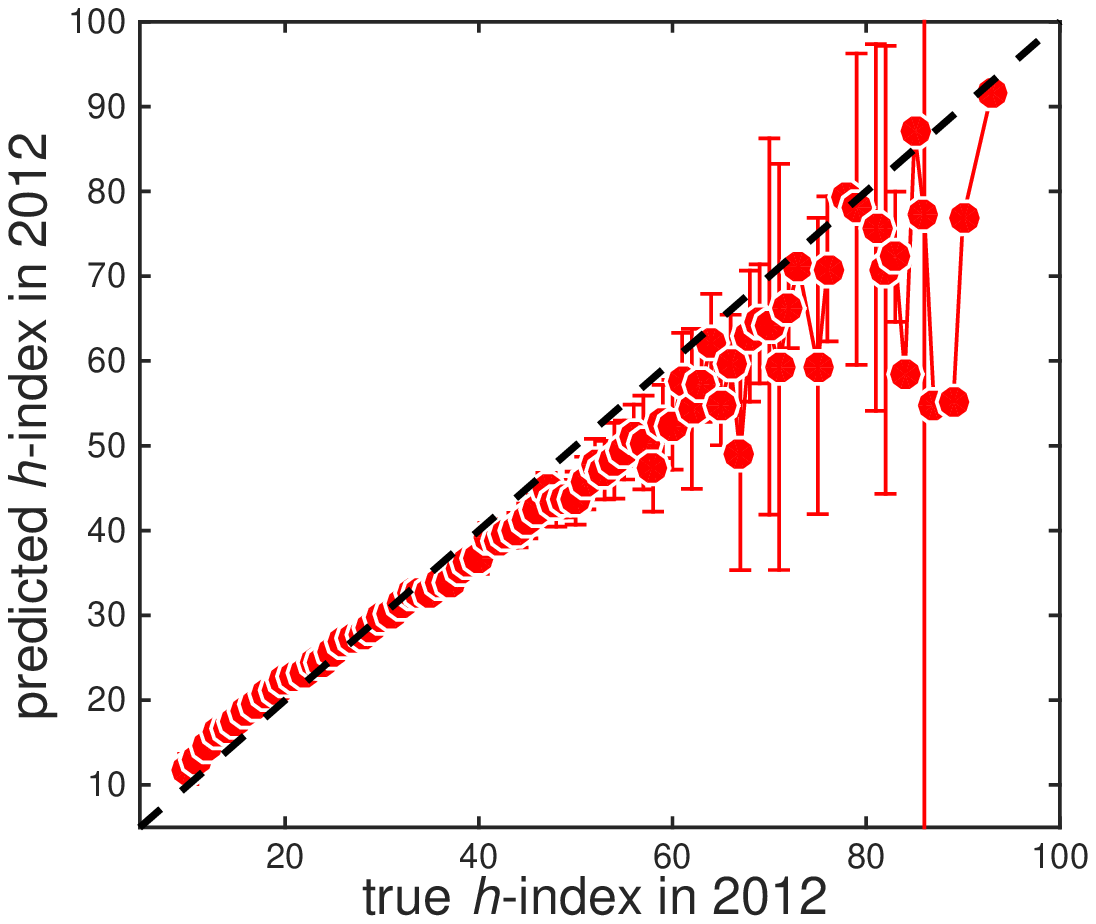}
}
\hspace{-0.1in}
\subfigure[\scriptsize Prediction from 2007 to 2012]{
\label{figsub:pred-h-2007}
\includegraphics[width=1.75in]{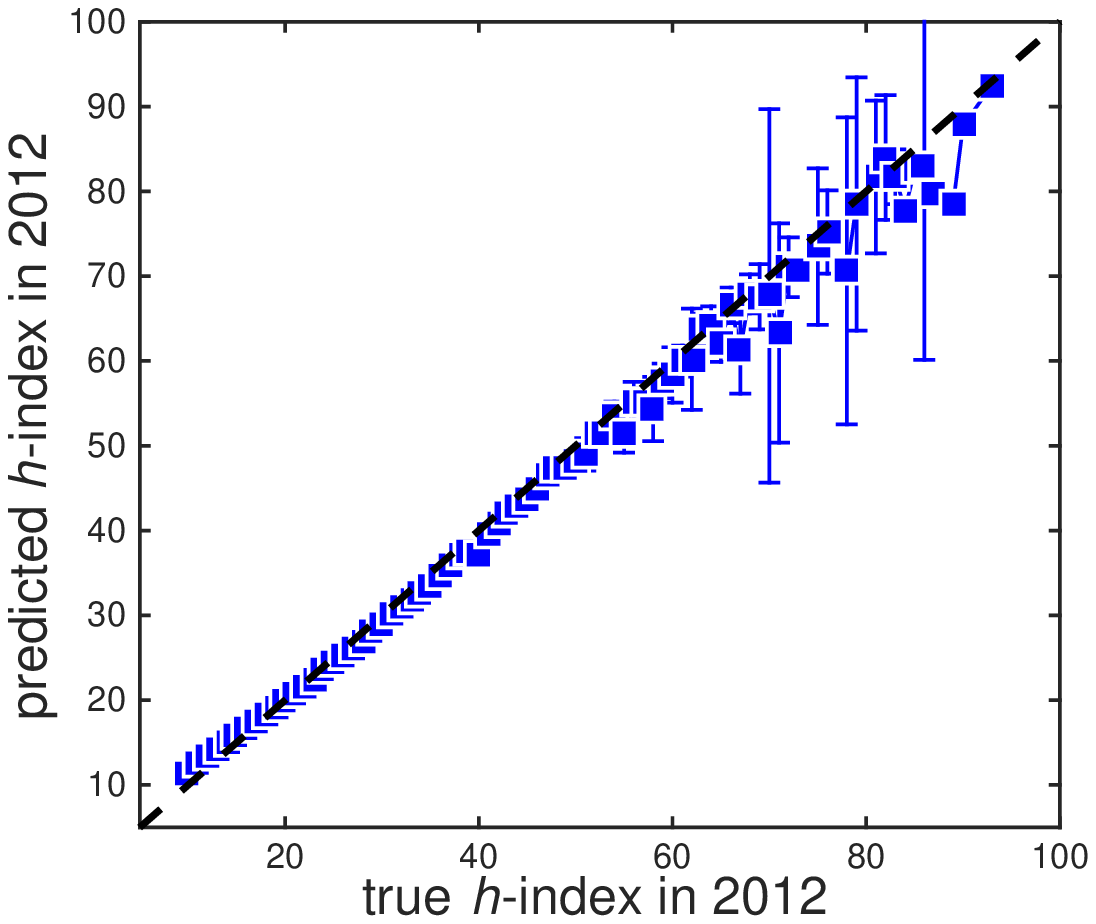}
\hspace{-0.2in}
}
\caption{\label{fig:hindex-pred}
{\bf \hindices\ in data vs. predicted \hindices.}
}
\vspace{-0.3cm}
\end{figure}

Figure~\ref{fig:hindex-pred} illustrates the concordance between the future \hindices\ predicted by our model and the actual \hindices\ according to the provided data. 
As the prediction timeframe can be varied, Figure~\ref{figsub:pred-h-2002} reports results over a ten-year timeframe, while Figure~\ref{figsub:pred-h-2007} reports results over a five-year timeframe. 
For both plots, optimal performance is denoted by the dashed $y=x$ line, which represents perfect agreement between the predictions and data. 
From the plots we observe that higher \hindices\ correspond to higher variability (error bars) and increasing deviation from optimal performance, suggesting that higher future \hindices\ are more difficult to predict. 
However, Figure~\ref{figsub:pred-h-2002} also demonstrates higher levels of deviation and variability than Figure~\ref{figsub:pred-h-2007}, indicating that accurately predicting future \hindices\ is more difficult over longer timeframes.

\subsection{Predicting Whether Papers Increase \hindices}

\vpara{Methods.}
Our problem of predicting whether a paper increases its primary author's future \hindex\ is formulated as a classification task. 
For this task, we employ a series of standard classification models, including logistic regression (LRC), support vector machine (SVM), na\"{\i}ve Bayes (NB), radial basis function network (RBF), bagged decision trees (BAG), and random forest (RF). 
Generally, we report the prediction results of each method to demonstrate the predictability of scientific impact, though we only use logistic regression to analyze factor contributions and parameter settings. 

Recall that for this task, we have defined two sets of papers, \textit{$P_{new}$} and \textit{$P_{old}$}, and we generate predictions for both. 
When defining the primary author as either the max-\hindex\ author or the first author, we further extract two sets of papers from both \textit{$P_{new}$} and \textit{$P_{old}$}, respectively, and have \textit{$P_{new}^{max}$}, \textit{$P_{new}^{first}$}, \textit{$P_{old}^{max}$}, and \textit{$P_{old}^{first}$}. 
For each set of papers, we use half of the instances (papers) in the set for model training and the remaining half for model validation. 
When predicting for $P_{new}$, we use the six groups of 24 total factors described in Table \ref{tb:factors}. 
When predicting for $P_{old}$, these 24 factors are used along with the three additional factors described in Table \ref{tb:existingfactors}. 
To quantitatively evaluate the predictability of the problem, we repeat the prediction experiments ten times and report the average performance in terms of precision, recall, \fonescore, area under the receiver operating characteristic (AUC), and accuracy. 
Furthermore, as our problem can be viewed as a ranking task (i.e., rank all of a scholar's papers in the reverse order of probability that they will increase her/his \hindex), the precision at the top 3 results (Pre@3) and mean average precision (MAP) are also used to evaluate performance.

\begin{table*}[t]
\caption{ {\bf Predictive performance for whether papers published at time $t$ (P$_{new}$) will increase their primary authors' future \hindex\ at $t$ + $\Delta t$.}  
\scriptsize{
The number in parentheses is the standard deviation. 
$t$=2007, $\Delta t$=5 years, and the \hindex\ threshold is set to 10. 
}
}
\label{tb:results-2007-new}
\small
\centering
\renewcommand\arraystretch{1.3}
\begin{tabular}{@{}l||c||l|l|l|l|l|l|l}

\hline
& Method  & Precision   & Recall          & F$_{1}$        &   AUC           &   Accuracy         &    Pre@3  & MAP \\
\hline \hline
\multirow{7}*{$P_{new}^{max}$} & Random  & 0.2107         & 0.5000          & 0.2965          & 0.5000         &   0.5000             & 0.5899    & 0.4132   \\
&LRC     & 0.8233 (0.0049)& 0.5929 (0.0062) & 0.6894 (0.0038) & 0.9299 (0.0017) & 0.8873 (0.0010)    & 0.8928    & 0.9440 \\
&SVM     & 0.8377 (0.0050)& 0.5806 (0.0044) & 0.6858 (0.0034) & 0.7753 (0.0021) & 0.8879 (0.0011)    & 0.8033    & 0.8655 \\
&NB      & 0.6483 (0.0113)& 0.5371 (0.0151) & 0.5873 (0.0072) & 0.8497 (0.0043) & 0.8409 (0.0024)    & 0.8201    & 0.8759 \\
&RBF     & 0.6679 (0.0109)& 0.5573 (0.0124) & 0.6075 (0.0081) & 0.8403 (0.0078) & 0.8482 (0.0029)    & 0.7897    & 0.8694 \\
&BAG     & 0.7992 (0.0045)& 0.7455 (0.0111) & 0.7713 (0.0043) & 0.9548 (0.0008) & 0.9068 (0.0009)    & 0.8919    & 0.9509 \\
&RF      & 0.7647 (0.0058)& 0.7630 (0.0090) & 0.7638 (0.0043) & 0.9373 (0.0015) & 0.9005 (0.0016)    & 0.8734    & 0.9376  \\
\hline \hline
\multirow{7}*{$P_{new}^{first}$} & Random  & 0.2660         & 0.5000          & 0.3472          & 0.5000         &   0.5000            & 0.8068    & 0.6728   \\
&LRC     & 0.8202 (0.0106)& 0.6129 (0.0131) & 0.7014 (0.0077) & 0.9112 (0.0028) & 0.8611 (0.0027)    & 0.9200    & 0.9647 \\
&SVM     & 0.7866 (0.0207)& 0.4893 (0.0134) & 0.6031 (0.0114) & 0.7205 (0.0065) & 0.8059 (0.0048)    & 0.8666    & 0.9094 \\
&NB      & 0.6776 (0.0149)& 0.5176 (0.0234) & 0.5865 (0.0143) & 0.8316 (0.0064) & 0.8130 (0.0046)    & 0.8733    & 0.9250 \\
&RBF     & 0.6895 (0.0167)& 0.5418 (0.0252) & 0.6064 (0.0163) & 0.8200 (0.0059) & 0.8661 (0.0057)    & 0.8866    & 0.9277 \\
&BAG     & 0.7815 (0.0103)& 0.6901 (0.0092) & 0.7329 (0.0068) & 0.9216 (0.0023) & 0.8661 (0.0035)    & 0.9000    & 0.9609 \\
&RF      & 0.7322 (0.0139)& 0.7136 (0.0131) & 0.7227 (0.0111) & 0.9033 (0.0034) & 0.8542 (0.0060)    & 0.8800    & 0.9518  \\
\hline
\end{tabular}
\vspace{-0.25cm}
\end{table*}

\begin{table*}[t]
\caption{{\bf Predictive performance for whether papers published before time $t$ (P$_{old}$) will increase their primary authors' future \hindex\ at $t$ + $\Delta t$.}  
\scriptsize{
The number in parentheses is the standard deviation. 
$t$=2007, $\Delta t$=5 years, and the \hindex\ threshold is set to 10. 
}
}
\label{tb:results-2007-old}
\small
\centering
\renewcommand\arraystretch{1.3}
\begin{tabular}{@{}l||c||l|l|l|l|l|l|l}

\hline
& Method     & Precision   & Recall          & F$_{1}$        &   AUC           &   Accuracy         &    Pre@3  & MAP \\
\hline \hline
\multirow{7}*{$P_{old}^{max}$} & Random  & 0.3776         & 0.5000          & 0.4303          & 0.5000         &   0.5000           & 0.5070          & 0.3186  \\
& LRC     & 0.9840 (0.0006)& 0.9829 (0.0008) & 0.9834 (0.0004) & 0.9995 (0.0000) & 0.9874 (0.0003)    & 0.9992    & 0.9992 \\
& SVM     & 0.9835 (0.0009)& 0.9806 (0.0014) & 0.9820 (0.0008) & 0.9853 (0.0007) & 0.9864 (0.0005)    & 0.9825    & 0.9844 \\
& NB      & 0.9316 (0.0024)& 0.8290 (0.0040) & 0.8773 (0.0022) & 0.9763 (0.0008) & 0.9124 (0.0014)    & 0.9620    & 0.9601 \\
& RBF     & 0.7860 (0.1066)& 0.6965 (0.1440) & 0.7211 (0.0533) & 0.8768 (0.0060) & 0.8019 (0.0181)    & 0.8933    & 0.8902 \\
& BAG     & 0.9939 (0.0005)& 0.9898 (0.0003) & 0.9918 (0.0003) & 0.9998 (0.0000) & 0.9938 (0.0002)    & 0.9998    & 0.9997 \\
& RF      & 0.9816 (0.0020)& 0.9880 (0.0003) & 0.9848 (0.0011) & 0.9992 (0.0001) & 0.9884 (0.0008)    & 0.9984    & 0.9984 \\
\hline \hline
\multirow{7}*{$P_{old}^{first}$} & Random  & 0.4753         & 0.5000          & 0.4873          & 0.5000         &   0.5000            & 0.6424    & 0.4524   \\
& LRC     & 0.9818 (0.0011)& 0.9803 (0.0007) & 0.9810 (0.0004) & 0.9988 (0.0000) & 0.9819 (0.0003)    & 0.9990    & 0.9994 \\
& SVM     & 0.9838 (0.0056)& 0.9725 (0.0085) & 0.9781 (0.0024) & 0.9790 (0.0024) & 0.9792 (0.0021)    & 0.9827    & 0.9865 \\
& NB      & 0.9588 (0.0030)& 0.7963 (0.0051) & 0.8700 (0.0024) & 0.9713 (0.0009) & 0.8868 (0.0017)    & 0.9740    & 0.9814 \\
& RBF     & 0.8956 (0.0244)& 0.4829 (0.0505) & 0.6259 (0.0428) & 0.8288 (0.0226) & 0.7271 (0.0218)    & 0.8810    & 0.8932 \\
& BAG     & 0.9873 (0.0010)& 0.9842 (0.0009) & 0.9858 (0.0004) & 0.9993 (0.0001) & 0.9865 (0.0003)    & 0.9990    & 0.9993 \\
& RF      & 0.9762 (0.0024)& 0.9828 (0.0009) & 0.9795 (0.0013) & 0.9982 (0.0002) & 0.9804 (0.0012)    & 0.9975    & 0.9985  \\
\hline
\end{tabular}
\vspace{-0.25cm}
\end{table*}

\vpara{Prediction Results for $P_{new}$.} 
The predictability of whether a paper published at $t$ = 2007 will contribute to its primary author's future \hindex\ within $\Delta t$ = 5 years is presented in Table~\ref{tb:results-2007-new}. 
The prediction is applied to the papers whose primary author had an \hindex\ in 2007 of at least 10. 
The resulting set when considering the max-\hindex\ author as the primary author, $P_{new}^{max}$, contains 29,254 papers, of which 21.07\% successfully contributed to their primary author's future \hindex\ by 2012. 
When the first author serves the primary author, the resulting set $P_{new}^{first}$ covers 9,231 papers, of which 26.60\% increased the first author's future \hindex\ by 2012. 
 
Overall, when predicting $P_{new}^{max}$, random guessing achieves an \fonescore\ of 0.2965 and an accuracy of 0.5000. 
However, our methodology achieves a predictive power that significantly outperforms random guessing, demonstrating an \fonescore\ ranging from 0.5873 to 0.7713 (+98\% to +160\% increase) and an accuracy ranging from 0.7753 to 0.9548 (+66\% to +91\%  increase). 
The performance is similarly promising when measured by precision, recall, and AUC. 
Furthermore, by ranking all of a scholar's publications in the reverse order of probability that they increase his or her \hindex, logistic regression can achieve a Pre@3 of 0.8928 and a MAP of 0.9440. 
Similarly, the experimental performance when predicting for $P_{new}^{first}$, where the first author is considered the primary author, significantly outperforms random guessing and demonstrates a comparable predictability with the results for $P_{new}^{max}$.

\vpara{Prediction Results for $P_{old}$.} 
The predictability of whether a paper published before $t = 2007$ will contribute to its primary author's future \hindex\ ($\geq$ 10) within $\Delta t = 5$ years is presented in Table~\ref{tb:results-2007-old}. 
The resulting set when considering the max-\hindex\ author (the first author) as the primary author, $P_{old}^{max}$ ($P_{old}^{first}$), contains 161,348 (85,704) papers, of which 37.76\% (47.53\%) successfully contributed to their primary authors' future \hindices\ by 2012. 
Random guessing achieves an \fonescore\ of 0.4303 (0.4873), an AUC of 0.5000 (0.5000), and a Pre@3 of 0.5070 (0.6424). 
Generally, the algorithms can achieve at least twice the performance of  random guessing, as measured by all of the evaluation metrics employed. 
The results demonstrate strong predictability for this scientific impact prediction task, with performance scores ranging from 0.98--0.99 as measured by precision, recall, \fonescore, AUC, accuracy, Pre@3, and MAP. 

As the selected algorithms achieve similarly effective results, we use logistic regression to examine the remaining experiments---primarily owing to its interpretability.

\subsection{Predictability of Difficult Papers}

Our experimental results provide evidence for the predictability of whether a newly or previously published paper will contribute to the \hindex\ of its primary author within five years.
Yet, two intuitive questions naturally arise concerning this predictability:
First, is a primary author with a high or low \hindex\ more predictable? 
Second, is a paper more predictable given a long or short prediction timeframe? 

\begin{figure*}[t]
\centering
\hspace{-0.1in}
\subfigure[\scriptsize Predicting for $P_{new}^{max}$]{
\hspace{-0.2in}
\label{figsub:varying-para-f1-new}
\includegraphics[width=1.75in]{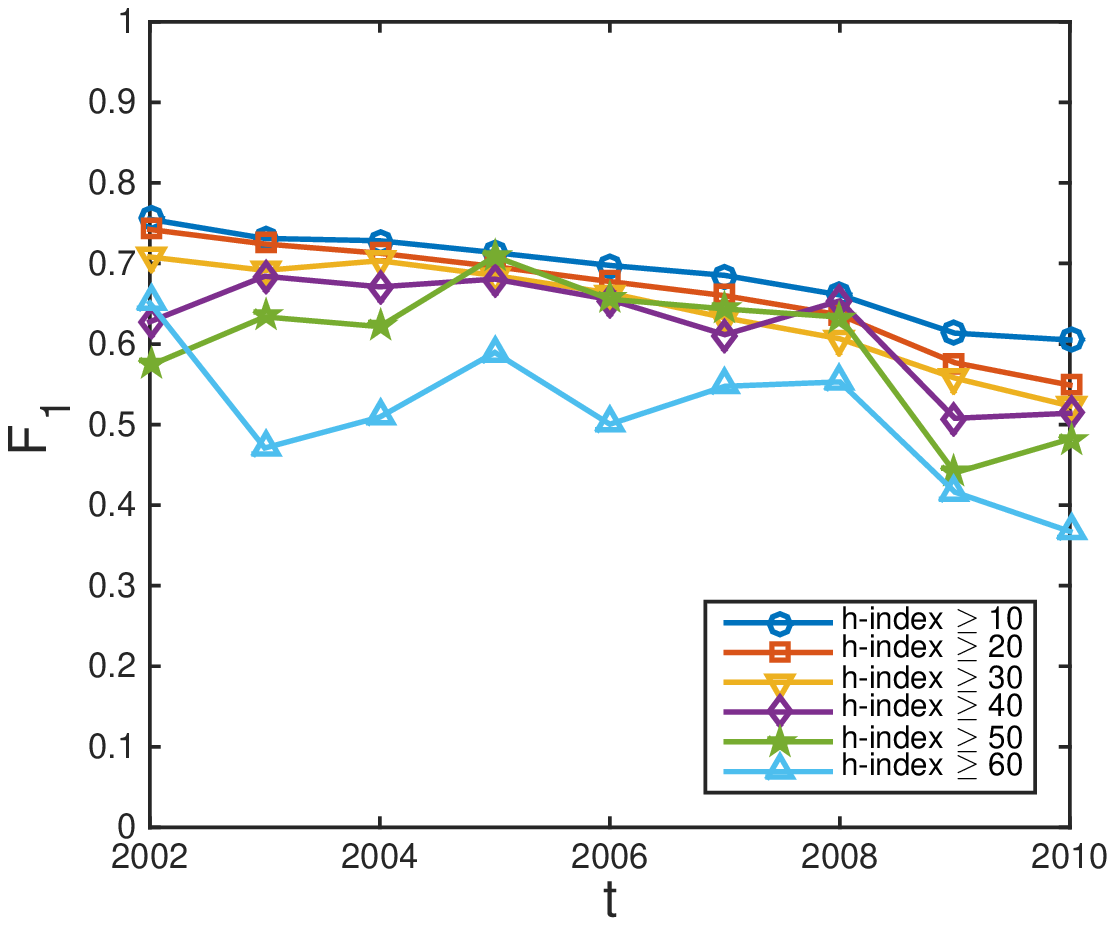}
}
\hspace{-0.1in}
\subfigure[\scriptsize Predicting for $P_{old}^{max}$]{
\label{figsub:varying-para-f1-old}
\includegraphics[width=1.75in]{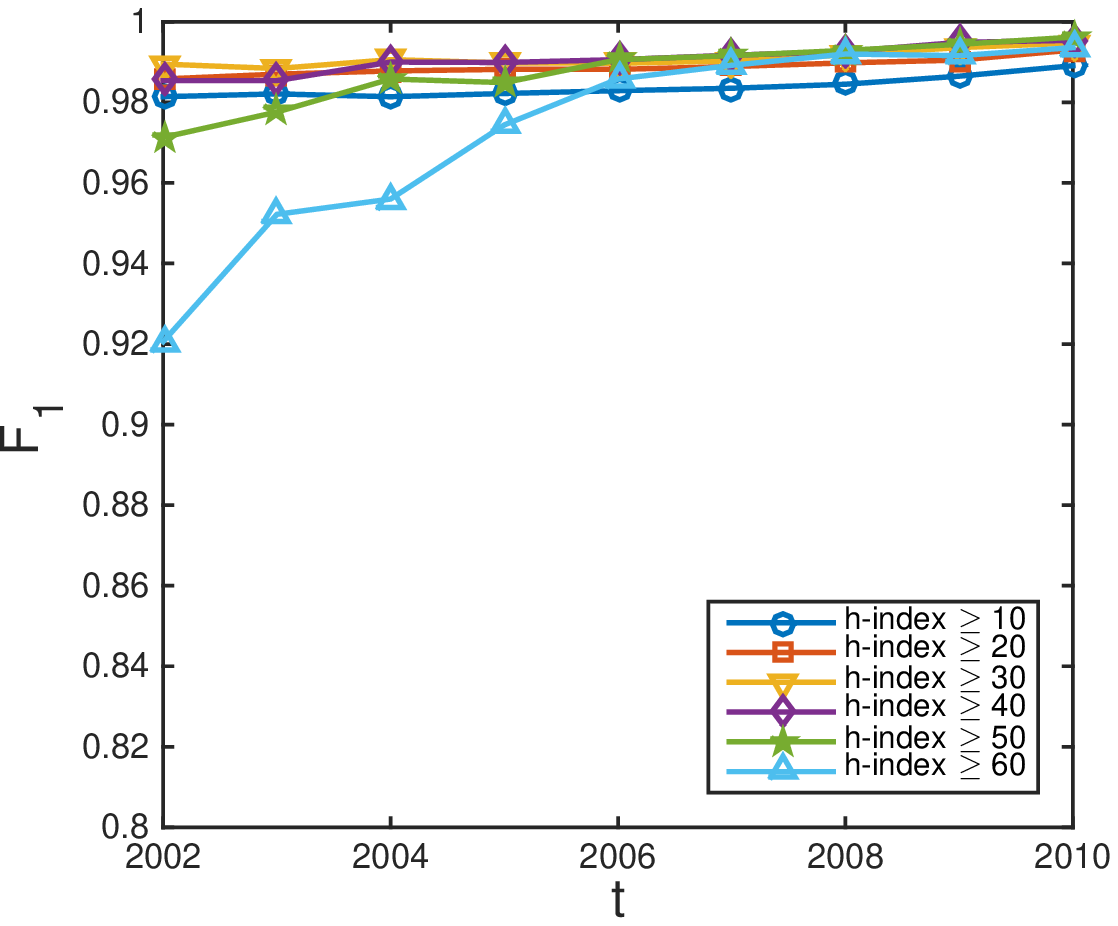}
\hspace{-0.2in}
}
\hspace{-0.1in}
\subfigure[\scriptsize Predicting for $P_{new}^{first}$]{
\label{figsub:varying-para-f1-new-first}
\includegraphics[width=1.75in]{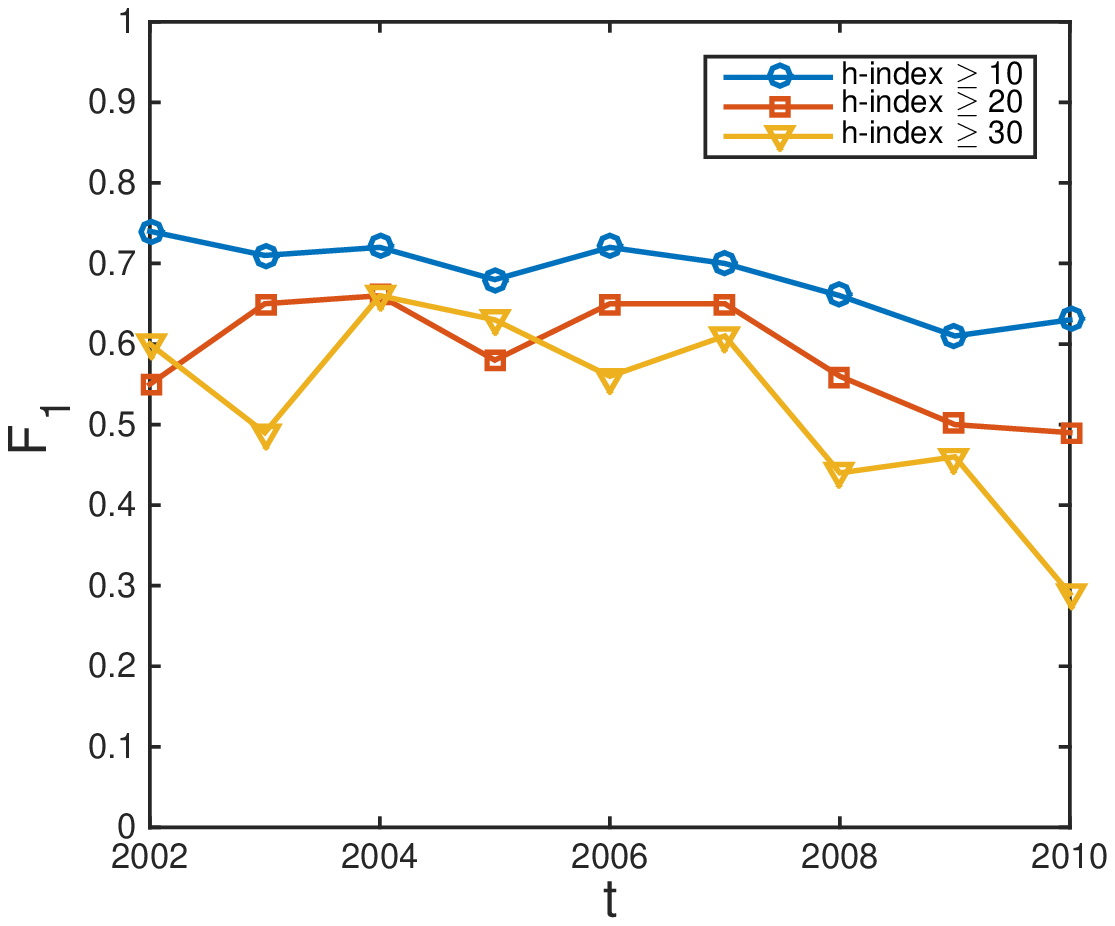}
\hspace{-0.2in}
}
\hspace{-0.1in}
\subfigure[\scriptsize Predicting for $P_{old}^{first}$]{
\label{figsub:varying-para-f1-old-first}
\includegraphics[width=1.75in]{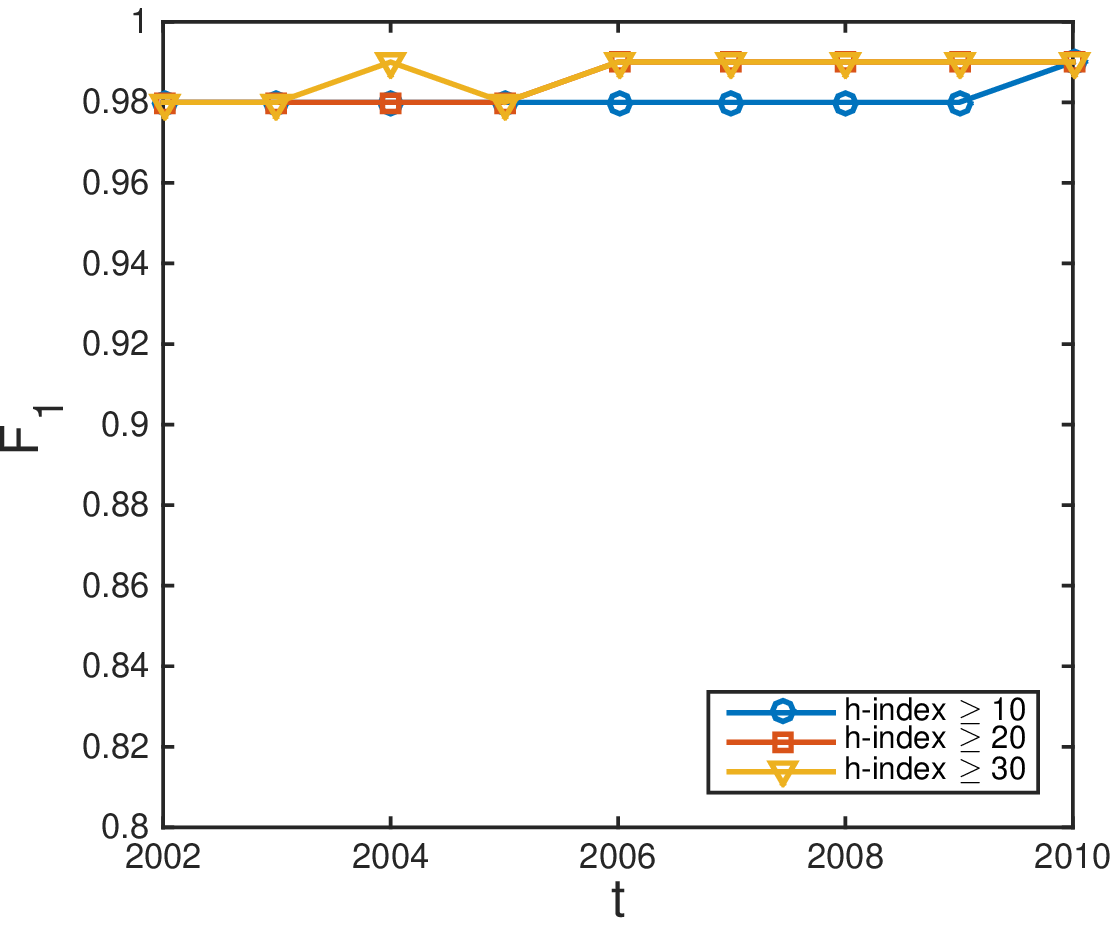}
\hspace{-0.2in}
}

\caption{\label{fig:varying-para}
{\bf Predictive performance for different papers.
} 
}
\vspace{-0.1cm}
\end{figure*}

To answer these questions, we investigate the predictability of papers conditioned on the primary author's \hindex\ and the length of the given prediction timeframe ($\Delta t$). 
Figure~\ref{fig:varying-para} shows the predictive performance given different constraints for four sets of papers, conditioned on the publication date and primary author definition---$P_{new}^{max}$, $P_{old}^{max}$, $P_{new}^{first}$, and $P_{old}^{first}$. 

First, from Figures~\ref{figsub:varying-para-f1-new} and \ref{figsub:varying-para-f1-new-first}, we find that predicting for papers with low-\hindex\ primary authors is a relatively easy task as measured by $F_1$ vis-\`a-vis predicting for those with high \hindices. 
 
Intuitively, the prediction task becomes increasingly non-trivial because of the increasing difficulty for any particular paper to reach the defined local threshold (i.e., the primary author's \hindex). 
Additionally, we observe that performance generally decreases as $t$ increases, implying that our prediction task is easier when given a longer timeframe $\Delta t=2012-t$. 
Intuitively, papers can accrue more citations as time goes on, during which time the authors' influence may increase, which may further compound the rate at which citations accrue. 
In summary, determining which newly published papers will increase one's \hindex\ is more predictable when conducted over a relatively long timeframe for those who have relatively low \hindices.

Note that from Figures~\ref{figsub:varying-para-f1-old} and \ref{figsub:varying-para-f1-old-first}, we can see that when predicting for previously published papers, both observations above are not significant. This is due to the relatively strong predictability of those papers.

\begin{figure*}[t]
\centering
\subfigure[\scriptsize $t$=2002, $P_{new}^{max}$]{
\hspace{-0.2in}
\label{figsub:fca-2002}
\includegraphics[width=1.75in]{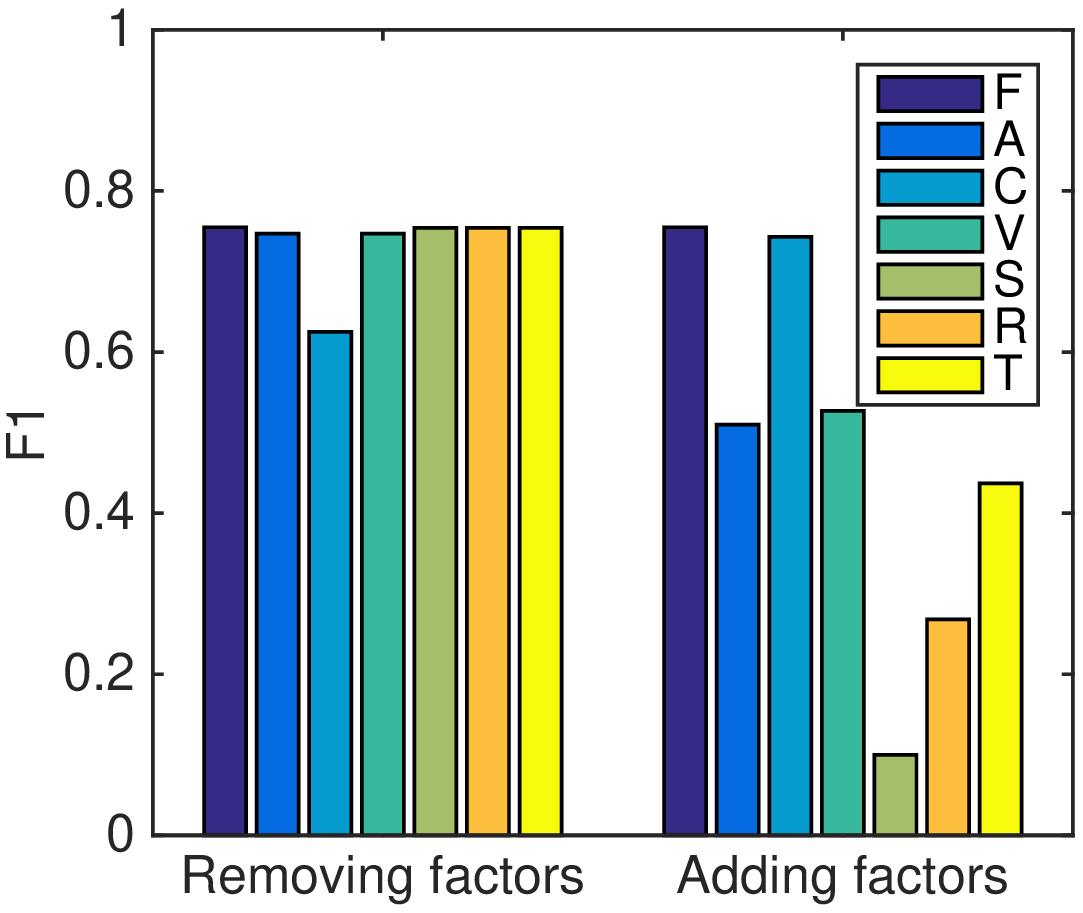}
}
\hspace{-0.1in}
\subfigure[\scriptsize $t$=2007, $P_{new}^{max}$]{
\label{figsub:fca-2007}
\includegraphics[width=1.75in]{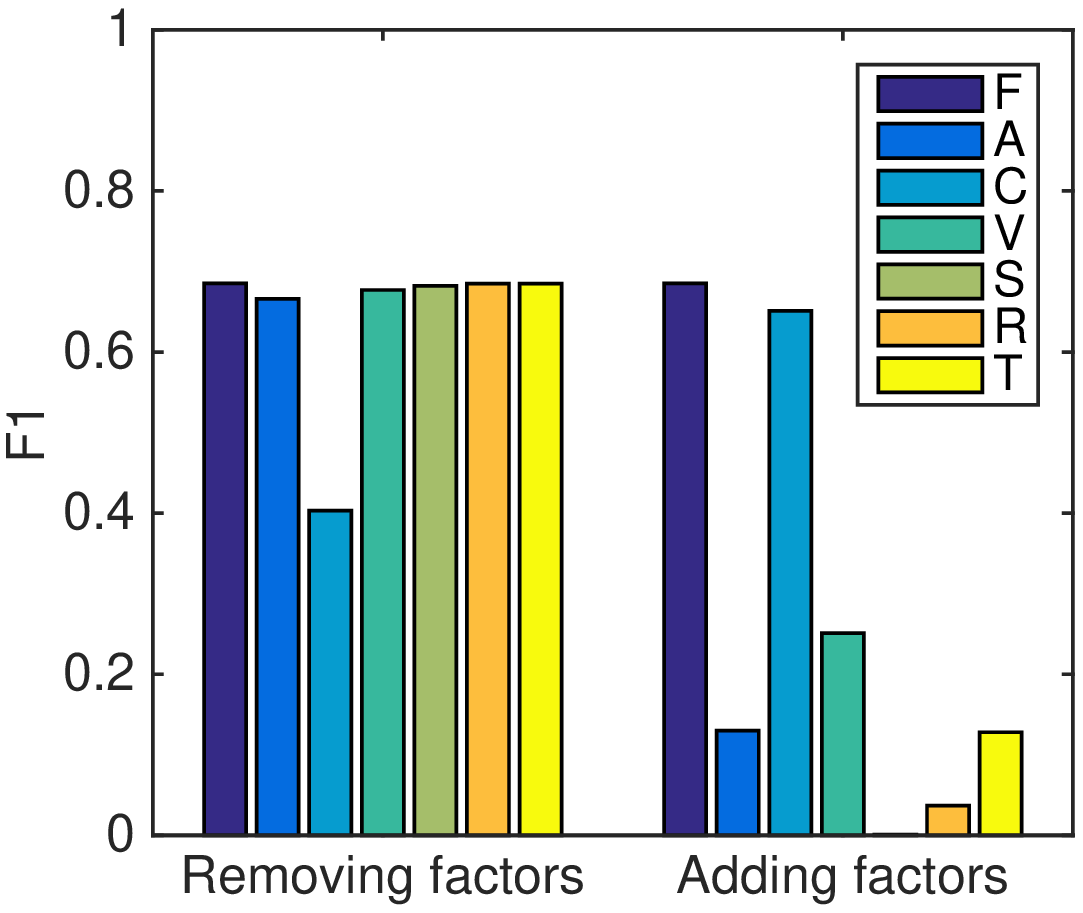}
\hspace{-0.2in}
}
\hspace{-0.1in}
\subfigure[\scriptsize $t$=2002, $P_{old}^{max}$]{
\label{figsub:fca-2002-old}
\includegraphics[width=1.75in]{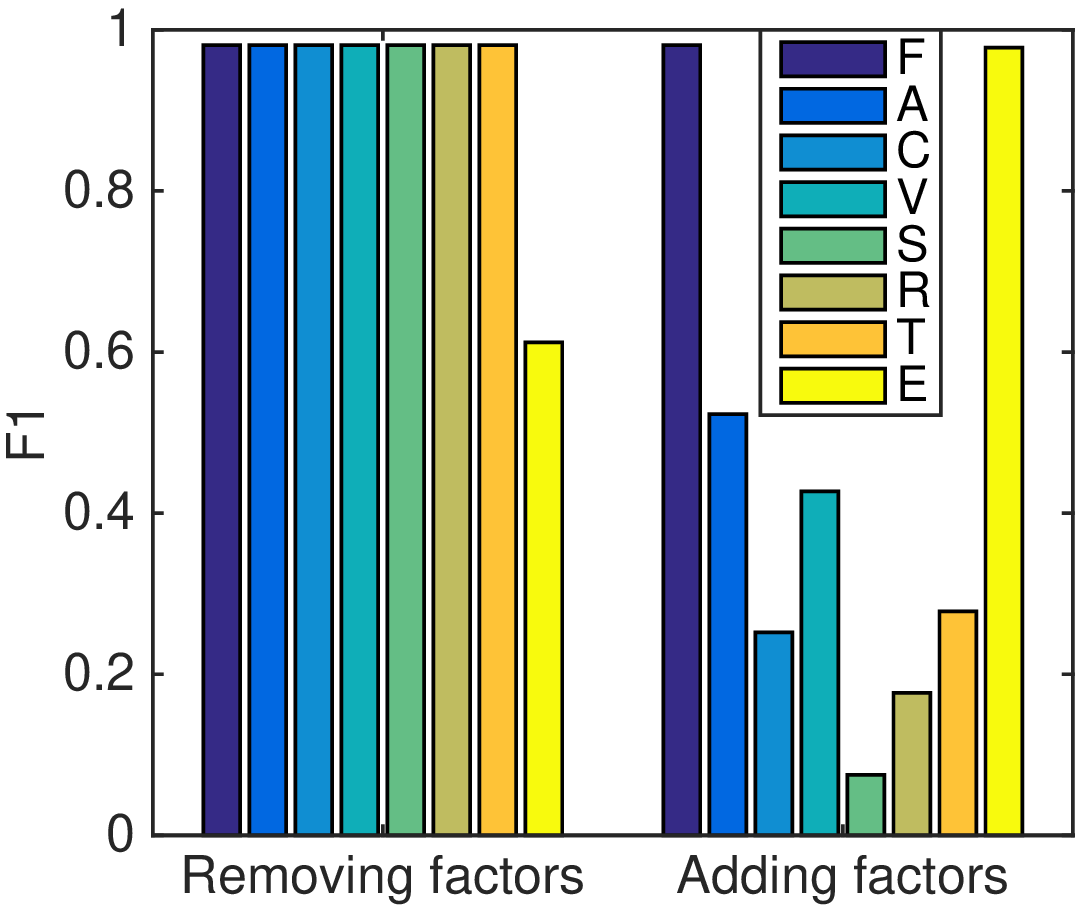}
\hspace{-0.2in}
}
\hspace{-0.1in}
\subfigure[\scriptsize $t$=2007, $P_{old}^{max}$]{
\label{figsub:fca-2007-old}
\includegraphics[width=1.75in]{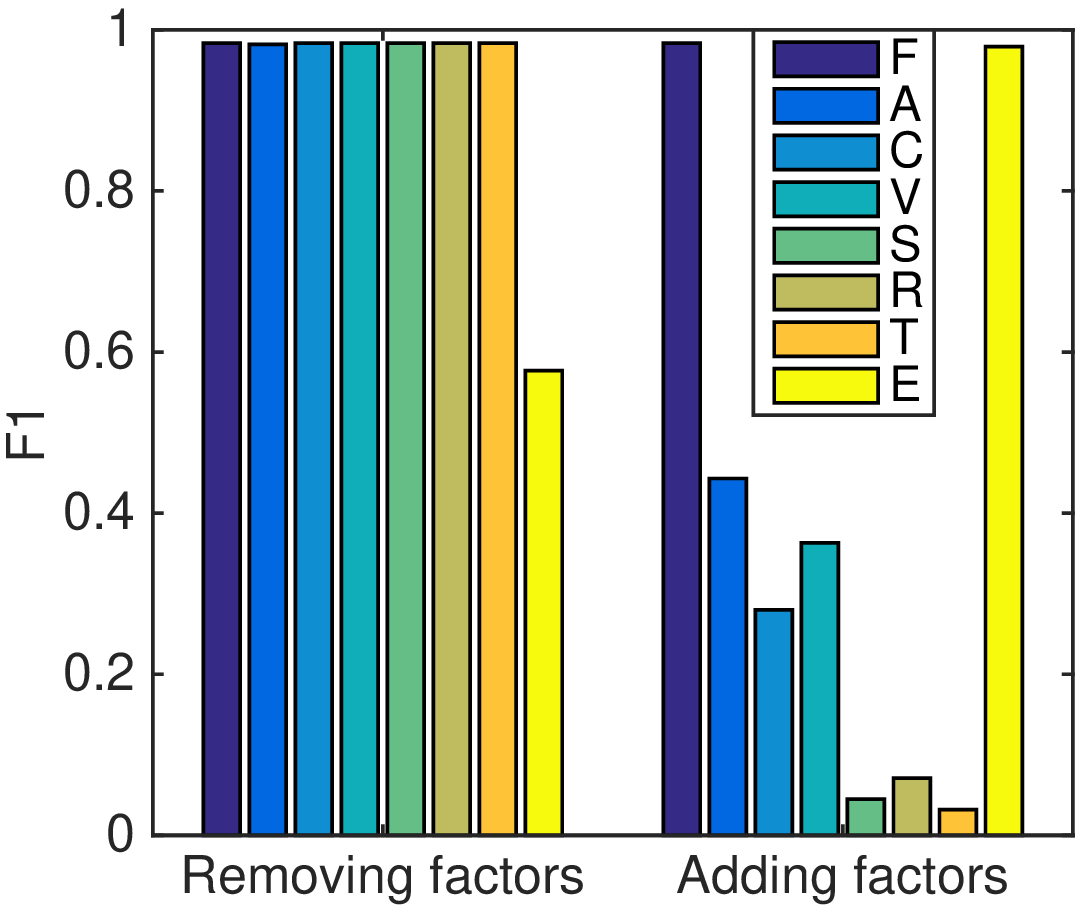}
\hspace{-0.2in}
}
\vspace{-0.25cm}
\subfigure[\scriptsize $t$=2002, $P_{new}^{first}$]{
\hspace{-0.2in}
\label{figsub:fca-2002-new-first}
\includegraphics[width=1.75in]{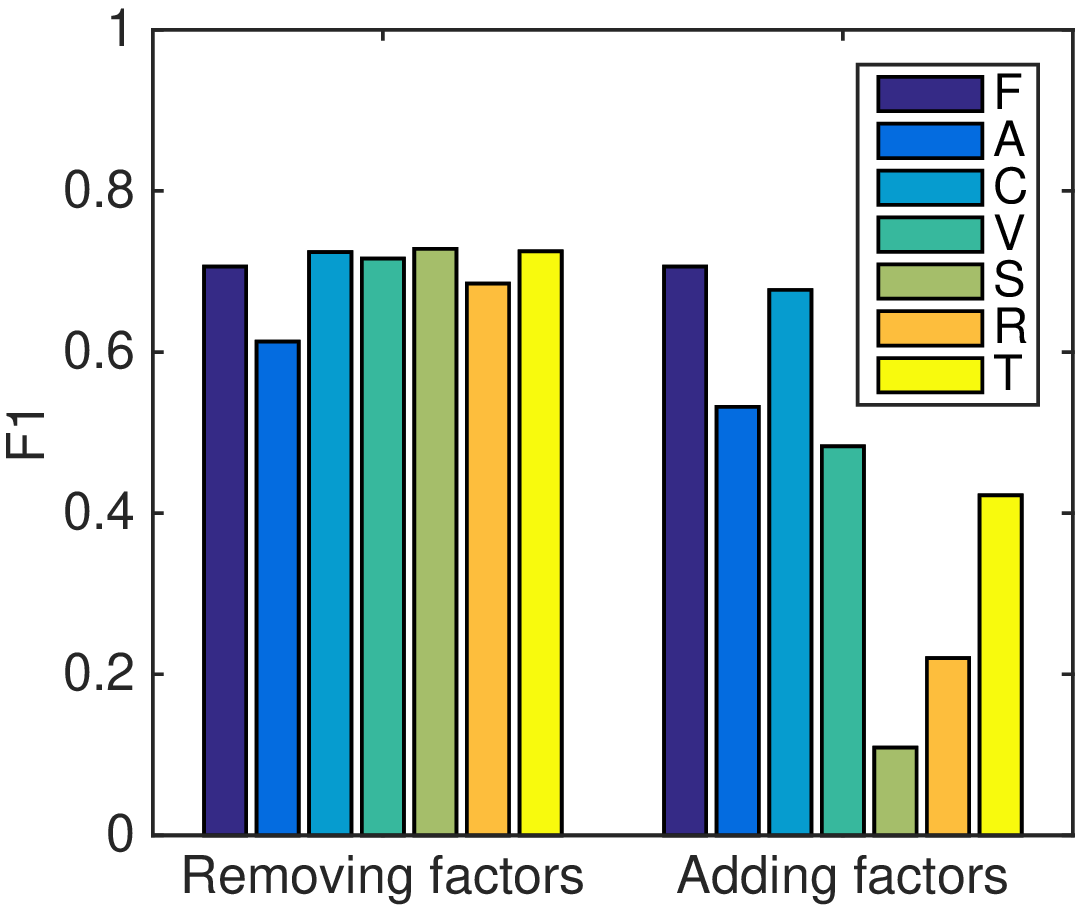}
}
\hspace{-0.1in}
\subfigure[\scriptsize $t$=2007, $P_{new}^{first}$]{
\label{figsub:fca-2007-new-first}
\includegraphics[width=1.75in]{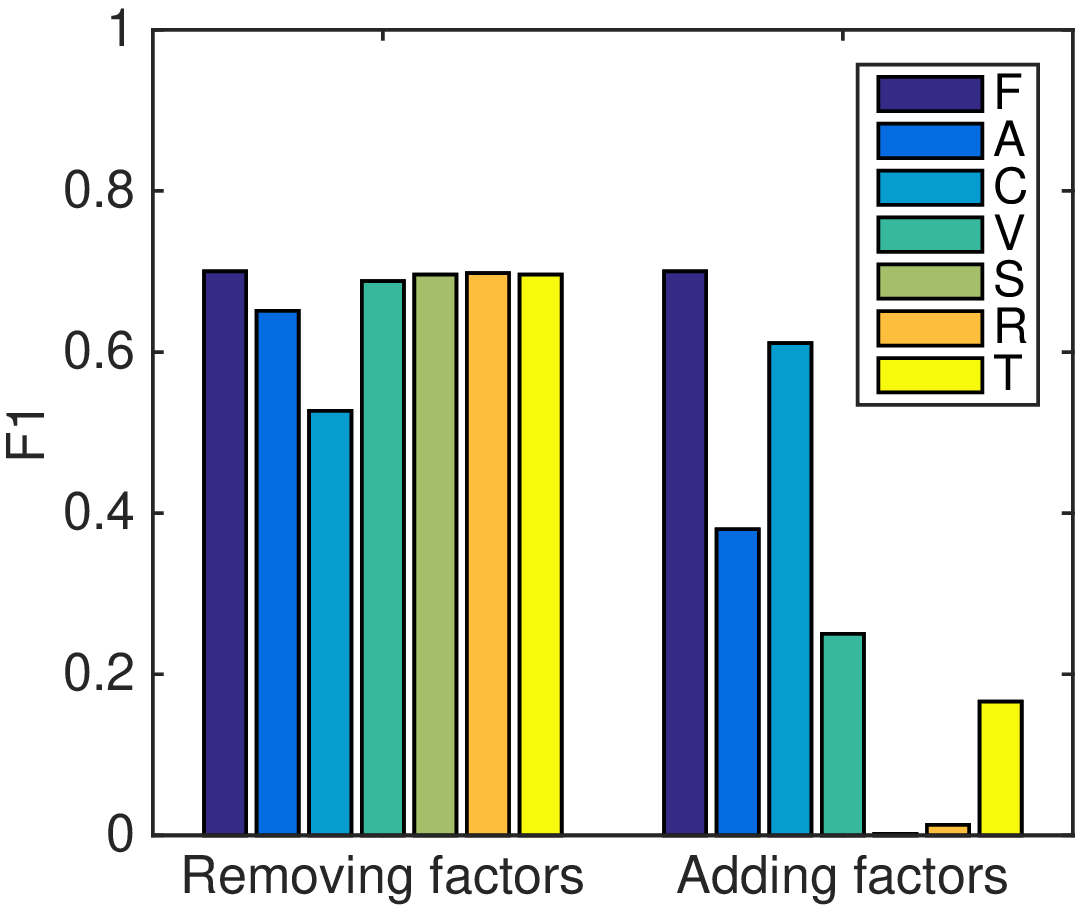}
\hspace{-0.2in}
}
\hspace{-0.1in}
\subfigure[\scriptsize $t$=2002, $P_{old}^{first}$]{
\label{figsub:fca-2002-old-first}
\includegraphics[width=1.75in]{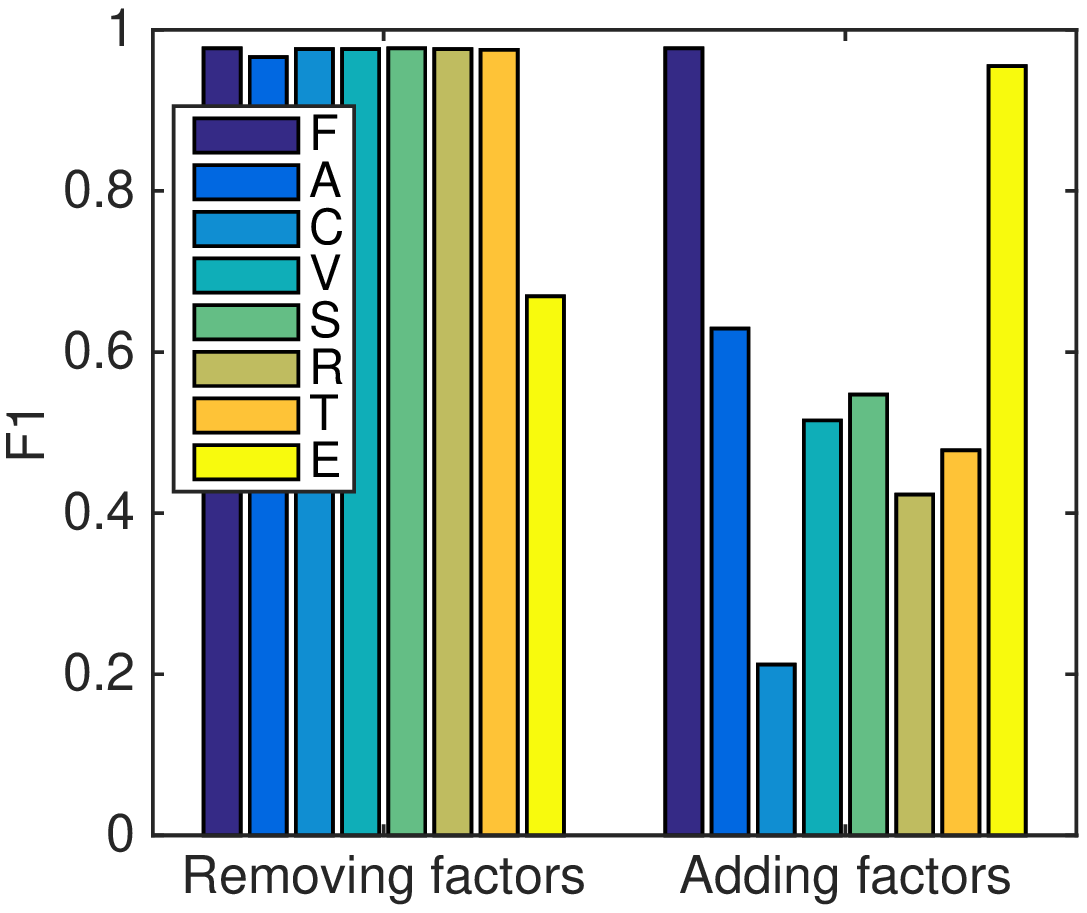}
\hspace{-0.2in}
}
\hspace{-0.1in}
\subfigure[\scriptsize $t$=2007, $P_{old}^{first}$]{
\label{figsub:fca-2007-old-first}
\includegraphics[width=1.75in]{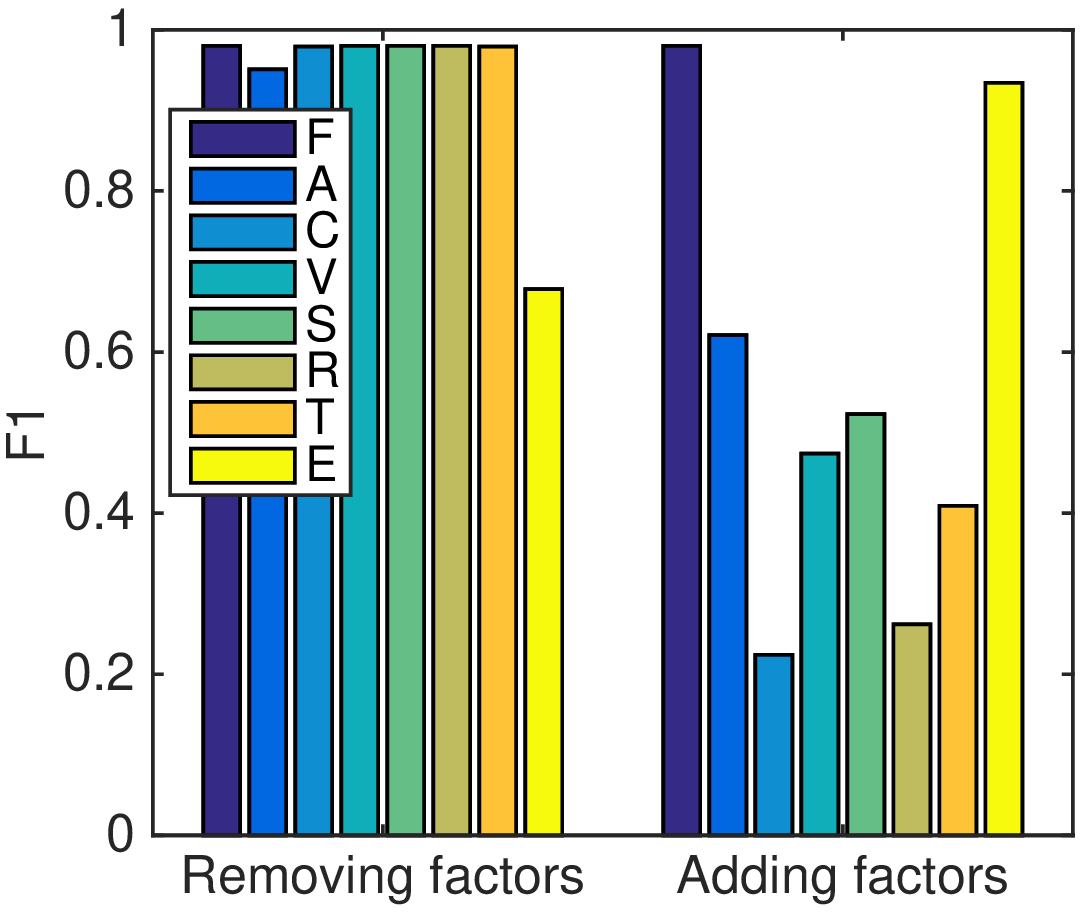}
\hspace{-0.2in}
}

\caption{\label{fig:factor-group-contribution}
{\bf Factor contribution analysis.} 
\scriptsize
Logistic regression model trained with only or without the denoted factors. F: full feature set; 
A: Author factors; C: Content factors; V: Venue factors; S: Social factors; R: Reference factors; T: Temporal factors; E: Existing factors for previously published papers. 
The left and right sides of the figure illustrate the effects of omitting (the ``without'' case) and only including (the ``with only'' case) the indicated group of factors for model training, respectively. 
}
\vspace{-0.2cm}
\end{figure*}

\subsection{Factor Contribution Analysis}

To predict whether a paper will increase its primary author's \hindex, we devise six diverse groups of factors (see \S\ref{sec:factor}) that may drive the growth of scientific impact. 

To explore the contributions and importance of each factor group to the prediction task, we employ a ``jackknife'' approach with two cases: (1) one at a time, we remove a group of factors and evaluate the predictive performance of our model trained only on the remaining five groups (the ``without'' case); and (2) one at a time, we use only a single group of factors and evaluate the predictive performance of our model trained only on this group (the ``with only'' case). This approach provides information on the individual contribution and unique information that each group of factors supplies to the overall prediction task. 
Figure~\ref{fig:factor-group-contribution} provides the \fonescore s for the two cases with different $t$ (2002 and 2007), primary authors (max-\hindex\ and first authors), and publication dates (new and old). 
We can see that the contributions of different groups of factors demonstrate a high degree of variability. 

In Figures \ref{figsub:fca-2002} and \ref{figsub:fca-2007}, the $\sim$20\% drop in \fonescore\ demonstrated by the removal of content factors indicates that they are critically important to predicting for $P_{new}^{max}$. 
By contrast, the marginal decreases in performance demonstrated by the removal of other types of factors imply that the remaining factors provide a limited amount of unique information. 
When used only by themselves, the content factors still play the most important role in predicting the growth of scientific impact, though venue factors also achieve a marked effect on performance. 
Furthermore, with the exclusion of content factors, all groups of factors demonstrate greater importance when employed over a longer timeframe $\Delta t$. 

From Figures \ref{figsub:fca-2002-old} and \ref{figsub:fca-2007-old}, we can see that the existing factors are crucially important to predicting for $P_{old}^{max}$, both by themselves (the ``with only'' contributions) and when used with other factors (the ``without'' contributions). 
Different from predicting for $P_{new}^{max}$ in Figures \ref{figsub:fca-2002} and \ref{figsub:fca-2007}, author factors play a more important role than both content and venue factors, observed from the ``with only'' factor contributions. 
Overall, we find that this contribution analysis is consistent with the factor correlation results elaborated upon in the previous section. 

Figures \ref{figsub:fca-2002-new-first} and \ref{figsub:fca-2007-new-first} show that when predicting for newly published papers, the content, author, and venue factors contribute the most to the increase of the first authors' future \hindices. 
Similarly, from Figures \ref{figsub:fca-2002-old-first} and \ref{figsub:fca-2007-old-first}, we can see that the existing information before $t$ is the most decisive factor group for predicting whether the previously published papers can contribute to the first authors' future \hindices. 
Surprisingly, we also find that different from the prediction cases in $P_{new}^{max}$, $P_{old}^{max}$, and $P_{new}^{first}$, the role of social factors is comparable with author and venue factors when predicting for $P_{old}^{first}$.

In summary, when predicting for the newly published papers in Figures \ref{figsub:fca-2002}, \ref{figsub:fca-2007}, \ref{figsub:fca-2002-new-first} and \ref{figsub:fca-2007-new-first}, the content factor group is most crucial to generating effective predictions, followed by venue, author, and temporal factors. 
However, observed from Figures \ref{figsub:fca-2002-old}, \ref{figsub:fca-2007-old}, \ref{figsub:fca-2002-old-first} and \ref{figsub:fca-2007-old-first}, the existing factor group is the most telling followed by author and venue factor groups when predicting for previously published papers. 
The group of content factors is important when predicting for the increase of the max-\hindex\ authors, while its effect is not significant compared to other factors when predicting the contribution to the first authors' \hindices.

We further examine the contributions of each individual factor to the prediction tasks. 
To assess each factor's importance, we employ the measure of information gain ratio (IGR) \cite{Kullback:51}, which is based on the expected reduction in entropy---that is, uncertainty---achieved by learning the state of a given factor. 
The higher the IGR for a given factor, the greater its measured importance.

Table~\ref{tb:factor-igr} lists the IGR and corresponding ranking for each individual factor. 
When considering the IGR for $P_{new}$, the factors that are indicative of an author's topical authority are the most important, including \textit{C-authority-max}, \textit{C-authority-ave} and \textit{C-authority-first}. 
Following in importance are the two venue factors. 
When considering the IGR for $P_{old}$, the factors that are indicative of the number of existing citations (\textit{E-numc} and \textit{E-numc-ave}) achieve the top two positions, followed by author authorities and venue factors. 
The IGR calculated for the remaining factors decreases to the next lowest order of magnitude, indicating that they provide relatively limited contributions to our prediction tasks.

\begin{table}[t]
\caption{{\bf Information gain ratio (IGR) of each factor.}
}
\scriptsize
\label{tb:factor-igr}
\centering
\renewcommand\arraystretch{1.3}
\begin{tabular}{l|l|l|l|l}

\hline

   \multirow{2}{*}{Factor} 
  &\multicolumn{2}{c|}{$P_{new}^{max}$}
  &\multicolumn{2}{|c}{$P_{old}^{max}$}\\
  \cline{2-5} & \scriptsize{IGR$_{2002}$ (R)}   &  \scriptsize{IGR$_{2007}$ (R)}  & \scriptsize{IGR$_{2002}$ (R)}         & \scriptsize{IGR$_{2007}$ (R)} \\ \hline
 
 \textit{A-first-max}     & 0.0193 (15) & 0.0255 (10) & 0.0168 (10) & 0.0206 (10)  \\ 
\textit{A-ave-max}                            & 0.0126 (19) & 0.0200 (11) & 0.0153 (11) & 0.0207 (9)\\ 
\textit{A-sum-max}                             & 0.0229 (13) & 0.0193 (12) & 0.0170 (9)  & 0.0134 (11)\\ 
\textit{A-first-ratio}                        & 0.0133 (17) & 0.0111 (15) & 0.0138 (12) & 0.0114 (12) \\ 
\textit{A-max-ratio}                           & 0.0631 (5)  & 0.0409 (7)  & 0.0665 (7)  & 0.0656 (7)\\ 
\textit{A-num-authors}                         & 0.0079 (20) & 0.0044 (23) & 0.0025 (21) & 0.0007 (26)\\ 

\hline

\textit{C-popularity}      & 0.0315 (11) & 0.0053 (20) & 0.0024 (23) & 0.0035 (23) \\  
\textit{C-diversity}                          & 0.0258 (12) & 0.0047 (22) & 0.0018 (26) & 0.0031 (25) \\ 
\textit{C-novelty}                            & 0.0127 (18) & 0.0062 (19) & 0.0018 (25) & 0.0000 (27)\\ 
\textit{C-auth.-first}                    & 0.3988 (1)  & 0.3407 (2)  & 0.0858 (3)  & 0.1269 (4)\\ 
\textit{C-auth.-max}                      & 0.3006 (3)  & 0.2651 (3)  & 0.0678 (6)  & 0.1081 (5)\\ 
\textit{C-auth.-ave}                      & 0.3781 (2)  & 0.3462 (1)  & 0.0854 (4)  & 0.1327 (3)\\ 

\hline

\textit{V-\hindex}      & 0.0619 (6) & 0.0714 (5) & 0.0494 (8) & 0.0586 (8)\\  
\textit{V-citation}                        & 0.1233 (4) & 0.1090 (4) & 0.0845 (5) & 0.1009 (6)\\  

\hline

\textit{S-degree}     & 0.0000 (24) & 0.0029 (24) & 0.0018 (24) & 0.0071 (19) \\ 
\textit{S-pagerank}   					  & 0.0000 (23) & 0.0052 (21) & 0.0025 (22) & 0.0089 (16)\\ 
\textit{S-h-coauthor} 					  & 0.0065 (21) & 0.0091 (17) & 0.0077 (21) & 0.0076 (17)\\ 
\textit{S-h-weight}   					  & 0.0045 (22) & 0.0078 (18) & 0.0051 (20) & 0.0056 (21)\\ 

\hline

\textit{R-\hindex}          & 0.0180 (16) & 0.0167 (14) & 0.0104 (16) & 0.0111 (14)\\ 
\textit{R-citation}                            & 0.0196 (14) & 0.0096 (16) & 0.0110 (14) & 0.0113 (13)\\ 

\hline

\textit{T-ave-h}       & 0.0551 (7)  & 0.0506 (6) & 0.0104 (17) & 0.0058 (20)\\  
\textit{T-max-h}   					   & 0.0476 (8)  & 0.0291 (9) & 0.0113 (13) & 0.0041 (22)\\ 
\textit{T-h-first} 					   & 0.0370 (9)  & 0.0386 (8) & 0.0108 (15) & 0.0072 (18) \\ 
\textit{T-h-max}   					   & 0.0341 (10) & 0.0168 (13)& 0.0093 (18) & 0.0034 (24) \\ 

\hline

\textit{E-numc}        &$\setminus$ & $\setminus$ & 0.7324 (2) & 0.7598 (1)\\  
\textit{E-numc-ave} 					   &$\setminus$ & $\setminus$ & 0.7336 (1) & 0.6477 (2)\\ 
\textit{E-num-years}  					   &$\setminus$ & $\setminus$ & 0.0002 (27)& 0.0105 (15) \\ 

\hline

\end{tabular}
\end{table}

\begin{figure*}[t]
\centering
\includegraphics[width=7.1in]{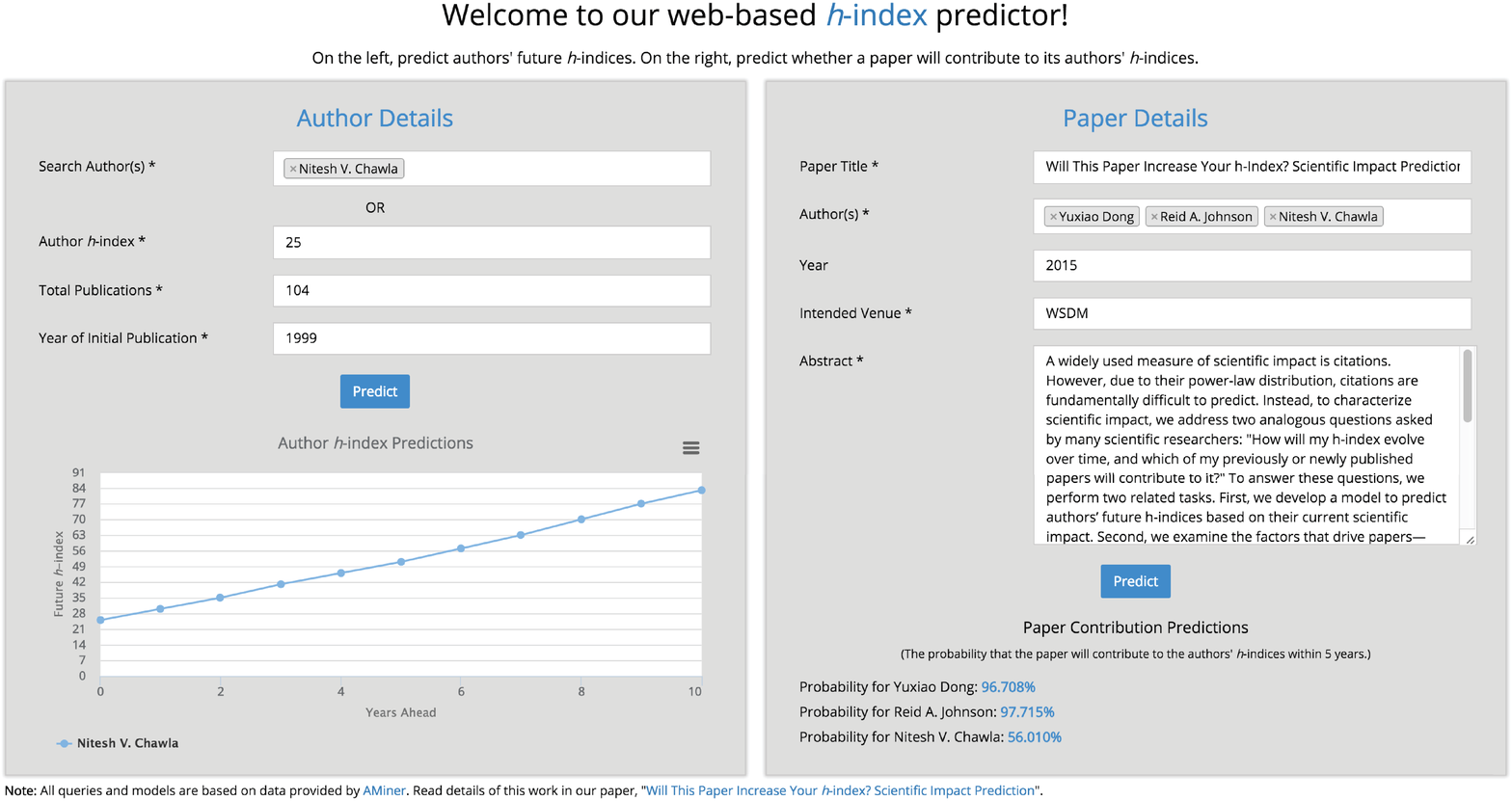}
\caption{\label{fig:demo}
{\bf Prototype \hindex\ prediction tool. } 
\scriptsize{
The prototype provides two distinct functionalities. On the left, the tool can be used to provide predictions of the development of authors' \hindices. On the right, the tool can be used to predict whether a paper will contribute to its authors' \hindices. 
}
}
\vspace{-0.2cm}
\end{figure*}

\subsection{Prototype \hindex\ Prediction Tool}
In light of our investigations into the factors that influence authors' \hindices, we have developed an online tool that allows users to generate \hindex\ predictions based on our findings\footnote{\scriptsize\url{http://www.icensa.com/hindex}}. 
An image of the working prototype is provided as Figure~\ref{fig:demo}. 

The tool provides separate functionality for predicting the development of authors' \hindices\ (left) and predicting whether a paper will contribute to its authors' \hindices\ (right). 
To predict the development of authors' \hindices, users may enter basic author details, such as an author's current \hindex, number of publications, and initial year of publication. 
To predict the probability that a paper will contribute to its authors' \hindices, users may enter basic paper details, such as the title, author list, year, venue, and abstract text. 
These details are then used to generate the factors described in this work, which serve as input to the \hindex\ growth or paper contribution model developed through our investigations.

We hope that the tool may be used by scholars to more effectively disseminate their work and to better gauge their future scientific impact.

\section{Related work}
\label{sec:related}

Scientific impact modeling is being extensively explored and has become an important and popular research topic~\cite{Acuna:Nature12,Uzzi:Science13,Wang:Science13,Ding:book14,Dong:ASONAM2015}. Its study offers the potential to help scholars more effectively disseminate their work and expand their scientific influence. 

Traditionally, the number of citations has been widely used as a measurement of scientific impact for both individual papers and solitary scientific researchers. 
Several practical metrics have been designed to reflect scientific impact based on citations. 
Garfield proposed the impact factor for indexing and evaluating the quality of journals~\cite{Garfield:Science1955}. 
More recently, Hirsch proposed the \hindex, which attempts to measure both a researcher's productivity and the popularity of his or her published work~\cite{Hirsch:05}. 
Both impact factor and \hindex\ successfully characterize the motivations and behavior of the scientific community, where scholars aspire to publish results in high-impact venues to increase their influence and \hindices\ and venues aim to publish cogent, influential work to improve their reputations and impact factors. 

Besides its measurement, a large body of work has been focused on the prediction of scientific impact. 
The 2003 ACM SIGKDD Cup introduced a competition focused around citation count prediction~\cite{KDDCUP:03}, with the task of estimating the number of times a paper has been cited given its previous number of citations. 
Following this, many efforts have been made to predict the number of future citations for scholarly work. 
Castillo et al. studied the correlation between author reputation and citations~\cite{Castillo:2007}. 
Yan et al. examined a series of features important to future citations~\cite{Yan:JCDL2012,Yan:CIKM2011}. 
Wang et al. uncovered basic mechanisms that govern scientific impact, which has the power to quantify and predict citation counts~\cite{Wang:Science13,Shen:AAAI14}. 
However, the effectiveness of such predictions is fundamentally limited by the heavy-tailed distribution of citations. 

Herein we (re)define the impact prediction problem by addressing a related question, namely: ``which of my papers will increase my (future) \hindex?'' 
The crucial difference between ours and previous work is that rather than trying to solve a regression task in a highly skewed environment, we instead tackle the problem by generating a local threshold (the author's \hindex) for each paper's future citation count.

Our work is also related to other mining tasks in academic data such as citation pattern and recommendation~\cite{Shi:JCDL2010,Ding:citation14,Ren:KDD14,Tang:2009citation,XiaoYu:SDM12}, topic influence~\cite{Liu:CIKM2010,Tang:09KDD}, information flow~\cite{Shi:PLOS2009,Shi:ICWSM2009}, collaboration prediction~\cite{Sun:WSDM2012,Wang:10KDD}, and analysis of citation networks~\cite{Smyth:ICML11} and academic social networks~\cite{Tang:08KDD}. 
Further, as the formalization of our predictive task is partly inspired by the cascade growth prediction problem~\cite{Cheng:WWW14}, the prediction of scientific impact is related to predicting the popularity~\cite{Hong:WSDM13,Pinto:WSDM13,Ahmed:WSDM13} of online ``paper'' (e.g., tweet, video, photo) in social media.

\section{Conclusion}
\label{sec:conclusion}

In this work, we study the predictability of scientific impact by formalizing two problems that can be reduced to the following questions: How will my \hindex\ evolve over time, and which of my papers will contribute to it? 
Our primary task is to determine whether a given paper, either previously or newly published, will increase the \textit{future} \hindex\ of its primary author within a predefined timeframe. 
To address this task, we first formalize an \hindex\ prediction problem to estimate researchers' future \hindices. 
We then use these estimates as the target for prediction in our primary task, which offers a powerful way of quantifying the interplay between researchers and publications and their effects on scientific impact. 

We find that two factors---topical authority and publication venue---are critical in determining whether a newly published paper will contribute to its primary author's future \hindex, while the existing citation count is the most decisive factor for a previously published paper. 
Surprisingly, we find that topic popularity and co-author influence have no statistical correlation with whether a paper will contribute to its primary author's future \hindex. 
We also find that the contribution of a paper to the impact of a researcher with a higher \hindex\ is generally more difficult to predict than for a researcher with a lower \hindex. 
Finally, we develop an \hindex\ prediction tool informed by our findings.  Overall, our work demonstrates a greater than 90\% potential predictability, as measured by accuracy, for whether a paper will contribute to its primary author's \hindex\ within five years.

Future work could study the interplay between a researcher's estimated future \hindex\ and the set of papers that we predict will contribute to his or her \hindex. 
Furthermore, as this work is conducted only on literature from computer science, examining other scientific disciplines for the same observed patterns could widen the scope and significance of our findings.

\scriptsize
\section*{Acknowledgments}
The authors would like to thank Mu Li at CMU for suggesting the \hindex\ demo. 
This work is supported by the Army Research Laboratory under Cooperative Agreement Number W911NF-09-2-0053, the U.S. Air Force Office of Scientific Research (AFOSR) and the Defense Advanced Research Projects Agency (DARPA) grant $\#$FA9550-12-1-0405, and the National Science Foundation (NSF) Grants OCI-1029584, BCS-1229450, and IIS-1447795.

\ifCLASSOPTIONcaptionsoff
  \newpage
\fi
\bibliographystyle{IEEEtran}
\bibliography{references-full}

\end{document}